\documentclass[reprint,twocolumn,floatfix,altaffilletter,superscriptaddress,preprintnumbers,
tightenlines,showpacs,showkeys,nofootinbib,notoccite,]{revtex4-2} 

\usepackage[utf8]{inputenc}
\usepackage[colorlinks=true,citecolor=blue,linkcolor=blue]{hyperref}
\usepackage[normalem]{ulem}
\usepackage{amsmath,amssymb, mathrsfs,esvect}
\usepackage{epsfig}
\usepackage{graphicx}               
\usepackage{url}
\usepackage{color}
\usepackage{slashed}
\usepackage{multirow}
\usepackage{placeins}
\usepackage[dvipsnames]{xcolor}
\usepackage{epstopdf}
\usepackage{soul}
\usepackage{tikz}
\usepackage[capitalise, english]{cleveref}
\usepackage{siunitx}
\usepackage{xspace}
\usepackage{booktabs}
\usepackage[capitalise]{cleveref}
\usepackage{dcolumn}
\usepackage{bm}
\usepackage{orcidlink}

\usetikzlibrary{trees}
\usetikzlibrary{decorations.pathmorphing}
\usetikzlibrary{decorations.markings}

\allowdisplaybreaks
\def\ie{{\frenchspacing\it i.e.}}

\def\be{\begin{equation}}
\def\ee{\end{equation}}
\def\ba{\begin{eqnarray}}
\def\ea{\end{eqnarray}}
\newcommand{\mockitem}[2]{%
  \noindent\textbf{#1}\quad #2\par
}

\begin{document}

\title{Nonlinear Information from DESI Luminous Red Galaxies: An Emulator-Based Analysis of Pre- and Post-Reconstruction Power Spectra}

\let\oldauthor\author
\RenewDocumentCommand{\author}{o m}{%
  \IfNoValueTF{#1}
    {\oldauthor{#2}}
    {\oldauthor{#2\,\orcidlink{#1}}}
}

\author[0000-0001-7756-8479]{Yuting Wang}
\email{ytwang@nao.cas.cn}
\affiliation{National Astronomical Observatories, Chinese Academy of Sciences, Beijing, 100101, P.R.China}

\author[0000-0003-4726-6714]{Gong-Bo Zhao}
\email{gbzhao@nao.cas.cn}
\affiliation{National Astronomical Observatories, Chinese Academy of Sciences, Beijing, 100101, P.R.China}
\affiliation{School of Astronomy and Space Science, University of Chinese Academy of Sciences, Beijing, 100049, P.R.China}

\author[0000-0001-6727-6915]{Kazuya Koyama}
\email{kazuya.koyama@port.ac.uk}
\affiliation{Institute of Cosmology and Gravitation, University of Portsmouth, Dennis Sciama Building, Portsmouth, PO1 3FX, UK}
\affiliation{Kavli IPMU (WPI), UTIAS, The University of Tokyo, Kashiwa, Chiba 277-8583, Japan}
\affiliation{Yukawa Institute for Theoretical Physics, Kyoto University, Kyoto 606-8502, Japan}

\author[0000-0002-7284-7265]{Ruiyang Zhao}
\affiliation{National Astronomical Observatories, Chinese Academy of Sciences, Beijing, 100101, P.R.China}

\author[0000-0002-9664-0760]{Takahiro Nishimichi}
\affiliation{Department of Astrophysics and Atmospheric Sciences, Faculty of Science, Kyoto Sangyo University, Motoyama, Kamigamo, Kitaku, Kyoto 603-8555, Japan}
\affiliation{Center for Gravitational Physics and Quantum Information, Yukawa Institute for Theoretical Physics, Kyoto University, Kyoto 606-8502, Japan}
\affiliation{Kavli Institute for the Physics and Mathematics of the Universe (WPI), The University of Tokyo Institutes for Advanced Study (UTIAS), The University of Tokyo, Chiba 277-8583, Japan}

\author[0000-0001-7984-5476]{Zhongxu Zhai}
\affiliation{Department of Astronomy, School of Physics and Astronomy, Shanghai Jiao Tong University, Shanghai 200240, China}
\affiliation{Shanghai Key Laboratory for Particle Physics and Cosmology, Shanghai 200240, China}

\author[0000-0003-0265-6217]{H\'ector Gil-Mar\'{\i}n}
\affiliation{ICC, University of Barcelona, IEEC-UB, Mart\'{\i} i Franqu\`es, 1, E-08028 Barcelona, Spain}

\author[0000-0001-6847-5254]{Hanyu Zhang}
\affiliation{Waterloo Centre for Astrophysics, University of Waterloo, 200 University Ave W, Waterloo, ON, N2L 3G1, Canada}
\affiliation{Department of Physics and Astronomy, University of Waterloo, 200 University Ave W, Waterloo, ON, N2L 3G1, Canada}

\author{Jessica Nicole Aguilar}
\affiliation{Lawrence Berkeley National Laboratory, 1 Cyclotron Road, Berkeley, CA 94720, USA}

\author[0000-0001-6098-7247]{Steven Ahlen}
\affiliation{Department of Physics, Boston University, 590 Commonwealth Avenue, Boston, MA 02215 USA}

\author[0000-0003-0467-5438]{Florian Beutler}
\affiliation{Institute for Astronomy, University of Edinburgh, Royal Observatory, Blackford Hill, Edinburgh EH9 3HJ, UK}

\author[0000-0001-9712-0006]{Davide Bianchi}
\affiliation{Dipartimento di Fisica ``Aldo Pontremoli'', Universit\`a degli Studi di Milano, Via Celoria 16, I-20133 Milano, Italy}
\affiliation{INAF-Osservatorio Astronomico di Brera, Via Brera 28, 20122 Milano, Italy}

\author{David Brooks}
\affiliation{Department of Physics \& Astronomy, University College London, Gower Street, London, WC1E 6BT, UK}

\author[0000-0001-7316-4573]{Francisco Javier Castander}
\affiliation{Institut d'Estudis Espacials de Catalunya (IEEC), c/ Esteve Terradas 1, Edifici RDIT, Campus PMT-UPC, 08860 Castelldefels, Spain}
\affiliation{Institute of Space Sciences, ICE-CSIC, Campus UAB, Carrer de Can Magrans s/n, 08913 Bellaterra, Barcelona, Spain}

\author{Todd Claybaugh}
\affiliation{Lawrence Berkeley National Laboratory, 1 Cyclotron Road, Berkeley, CA 94720, USA}

\author[0000-0002-2169-0595]{Andrei Cuceu}
\affiliation{Lawrence Berkeley National Laboratory, 1 Cyclotron Road, Berkeley, CA 94720, USA}

\author[0000-0002-1769-1640]{Axel de la Macorra}
\affiliation{Instituto de F\'{\i}sica, Universidad Nacional Aut\'{o}noma de M\'{e}xico,  Circuito de la Investigaci\'{o}n Cient\'{\i}fica, Ciudad Universitaria, Cd. de M\'{e}xico  C.~P.~04510,  M\'{e}xico}

\author[0000-0003-0920-2947]{Arnaud de Mattia}
\affiliation{IRFU, CEA, Universit\'en Paris-Saclay, F-91191 Gif-sur-Yvette, France}

\author[0000-0002-5665-7912]{Biprateep Dey}
\affiliation{Department of Astronomy \& Astrophysics, University of Toronto, Toronto, ON M5S 3H4, Canada}
\affiliation{Department of Physics \& Astronomy and Pittsburgh Particle Physics, Astrophysics, and Cosmology Center (PITT PACC), University of Pittsburgh, 3941 O'Hara Street, Pittsburgh, PA 15260, USA}

\author{Peter Doel}
\affiliation{Department of Physics \& Astronomy, University College London, Gower Street, London, WC1E 6BT, UK}

\author{Daniel J.~Eisenstein}
\affiliation{Center for Astrophysics | Harvard \& Smithsonian, 60 Garden Street, Cambridge, MA 02138, USA}

\author[0000-0003-4992-7854]{Simone Ferraro}
\affiliation{Lawrence Berkeley National Laboratory, 1 Cyclotron Road, Berkeley, CA 94720, USA}
\affiliation{University of California, Berkeley, 110 Sproul Hall \#5800 Berkeley, CA 94720, USA}

\author[0000-0002-2890-3725]{Jaime E.~Forero-Romero}
\affiliation{Departamento de F\'isica, Universidad de los Andes, Cra. 1 No. 18A-10, Edificio Ip, CP 111711, Bogot\'a, Colombia}
\affiliation{Observatorio Astron\'omico, Universidad de los Andes, Cra. 1 No. 18A-10, Edificio H, CP 111711 Bogot\'a, Colombia}

\author[0000-0001-9632-0815]{Enrique Gaztañaga}
\affiliation{Institut d'Estudis Espacials de Catalunya (IEEC), c/ Esteve Terradas 1, Edifici RDIT, Campus PMT-UPC, 08860 Castelldefels, Spain}
\affiliation{Institute of Cosmology and Gravitation, University of Portsmouth, Dennis Sciama Building, Portsmouth, PO1 3FX, UK}
\affiliation{Institute of Space Sciences, ICE-CSIC, Campus UAB, Carrer de Can Magrans s/n, 08913 Bellaterra, Barcelona, Spain}

\author[0000-0003-3142-233X]{Satya Gontcho A Gontcho}
\affiliation{University of Virginia, Department of Astronomy, Charlottesville, VA 22904, USA}

\author[0009-0007-9215-489X]{Gan Gu}
\affiliation{National Astronomical Observatories, Chinese Academy of Sciences, Beijing, 100101, P.R.China}
\affiliation{School of Astronomy and Space Science, University of Chinese Academy of Sciences, Beijing, 100049, P.R.China}

\author{Gaston Gutierrez}
\affiliation{Fermi National Accelerator Laboratory, PO Box 500, Batavia, IL 60510, USA}

\author[0000-0003-1197-0902]{ChangHoon Hahn}
\affiliation{University of Texas at Austin, Department of Astronomy, 2515 Speedway, Austin, TX 78712, USA}

\author[0000-0002-6550-2023]{Klaus Honscheid}
\affiliation{Center for Cosmology and AstroParticle Physics, The Ohio State University, 191 West Woodruff Avenue, Columbus, OH 43210, USA}
\affiliation{Department of Physics, The Ohio State University, 191 West Woodruff Avenue, Columbus, OH 43210, USA}
\affiliation{The Ohio State University, Columbus, 43210 OH, USA}

\author[0000-0002-1081-9410]{Cullan Howlett}
\affiliation{School of Mathematics and Physics, University of Queensland, Brisbane, QLD 4072, Australia}

\author[0000-0003-0201-5241]{Dick Joyce}
\affiliation{NSF NOIRLab, 950 N. Cherry Ave., Tucson, AZ 85719, USA}

\author[0000-0002-0000-2394]{Stephanie Juneau}
\affiliation{NSF NOIRLab, 950 N. Cherry Ave., Tucson, AZ 85719, USA}

\author{Robert Kehoe}
\affiliation{Department of Physics, Southern Methodist University, 3215 Daniel Avenue, Dallas, TX 75275, USA}

\author[0000-0002-8828-5463]{David Kirkby}
\affiliation{Department of Physics and Astronomy, University of California, Irvine, 92697, USA}

\author[0000-0003-3510-7134]{Theodore Kisner}
\affiliation{Lawrence Berkeley National Laboratory, 1 Cyclotron Road, Berkeley, CA 94720, USA}

\author{Jean-Paul Kneib}
\affiliation{Laboratory of Astrophysics, \`Ecole Polytechnique F\`ed\`erale de Lausanne (EPFL), Observatoire de Sauverny, CH-1290 Versoix, Switzerland}

\author[0000-0001-6356-7424]{Anthony Kremin}
\affiliation{Lawrence Berkeley National Laboratory, 1 Cyclotron Road, Berkeley, CA 94720, USA}

\author[0000-0002-6731-9329]{Claire Lamman}
\affiliation{The Ohio State University, Columbus, 43210 OH, USA}

\author[0000-0003-1838-8528]{Martin Landriau}
\affiliation{Lawrence Berkeley National Laboratory, 1 Cyclotron Road, Berkeley, CA 94720, USA}

\author[0000-0001-7178-8868]{Laurent Le Guillou}
\affiliation{Sorbonne Universit\'{e}, CNRS/IN2P3, Laboratoire de Physique Nucl\'{e}aire et de Hautes Energies (LPNHE), FR-75005 Paris, France}

\author[0000-0003-4962-8934]{Marc Manera}
\affiliation{Departament de F\'{i}sica, Serra H\'{u}nter, Universitat Aut\`{o}noma de Barcelona, 08193 Bellaterra (Barcelona), Spain}
\affiliation{Institut de F\'{i}sica d’Altes Energies (IFAE), The Barcelona Institute of Science and Technology, Edifici Cn, Campus UAB, 08193, Bellaterra (Barcelona), Spain}

\author[0000-0002-1125-7384]{Aaron Meisner}
\affiliation{NSF NOIRLab, 950 N. Cherry Ave., Tucson, AZ 85719, USA}

\author{Roman Miquel}
\affiliation{Instituci\'{o} Catalana de Recerca i Estudis Avan\c{c}ats, Passeig de Llu\'{\i}s Companys, 23, 08010 Barcelona, Spain}
\affiliation{Institut de F\'{i}sica d’Altes Energies (IFAE), The Barcelona Institute of Science and Technology, Edifici Cn, Campus UAB, 08193, Bellaterra (Barcelona), Spain}

\author[0000-0001-9070-3102]{Seshadri Nadathur}
\affiliation{Institute of Cosmology and Gravitation, University of Portsmouth, Dennis Sciama Building, Portsmouth, PO1 3FX, UK}

\author[0000-0001-8684-2222]{Jeffrey A.~Newman}
\affiliation{Department of Physics \& Astronomy and Pittsburgh Particle Physics, Astrophysics, and Cosmology Center (PITT PACC), University of Pittsburgh, 3941 O'Hara Street, Pittsburgh, PA 15260, USA}

\author[0000-0002-4637-2868]{Enrique Paillas}
\affiliation{Steward Observatory, University of Arizona, 933 N, Cherry Ave, Tucson, AZ 85721, USA}
\affiliation{Instituto de Estudios Astrof\'isicos, Facultad de Ingenier\'ia y Ciencias, Universidad Diego Portales, Av. Ej\'ercito Libertador 441, Santiago, Chile}

\author[0000-0002-0644-5727]{Will J.~Percival}
\affiliation{Department of Physics and Astronomy, University of Waterloo, 200 University Ave W, Waterloo, ON N2L 3G1, Canada}
\affiliation{Perimeter Institute for Theoretical Physics, 31 Caroline St. North, Waterloo, ON N2L 2Y5, Canada}
\affiliation{Waterloo Centre for Astrophysics, University of Waterloo, 200 University Ave W, Waterloo, ON N2L 3G1, Canada}

\author[0000-0001-7145-8674]{Francisco Prada}
\affiliation{Instituto de Astrof\'{i}sica de Andaluc\'{i}a (CSIC), Glorieta de la Astronom\'{i}a, s/n, E-18008 Granada, Spain}

\author[0000-0001-6979-0125]{Ignasi P\'erez-R\`afols}
\affiliation{Departament de F\'isica, EEBE, Universitat Polit\`ecnica de Catalunya, c/Eduard Maristany 10, 08930 Barcelona, Spain}

\author[0000-0001-7545-3504]{Alberto J.~Rosado-Mar\'{i}n}
\affiliation{Department of Physics \& Astronomy, Ohio University, 139 University Terrace, Athens, OH 45701, USA}

\author[0000-0002-7522-9083]{Ashley J.~Ross}
\affiliation{Center for Cosmology and AstroParticle Physics, The Ohio State University, 191 West Woodruff Avenue, Columbus, OH 43210, USA}
\affiliation{Department of Astronomy, The Ohio State University, 4055 McPherson Laboratory, 140 West 18th Avenue, Columbus, OH 43210, USA}
\affiliation{The Ohio State University, Columbus, 43210 OH, USA}

\author{Graziano Rossi}
\affiliation{Department of Physics and Astronomy, Sejong University, 209 Neungdong-ro, Gwangjin-gu, Seoul 05006, Republic of Korea}

\author[0000-0002-1609-5687]{Lado Samushia}
\affiliation{Abastumani Astrophysical Observatory, Tbilisi, GE-0179, Georgia}
\affiliation{Department of Physics, Kansas State University, 116 Cardwell Hall, Manhattan, KS 66506, USA}
\affiliation{Faculty of Natural Sciences and Medicine, Ilia State University, 0194 Tbilisi, Georgia}

\author[0000-0002-9646-8198]{Eusebio Sanchez}
\affiliation{CIEMAT, Avenida Complutense 40, E-28040 Madrid, Spain}

\author[0000-0002-3569-7421]{Edward F.~Schlafly}
\affiliation{Space Telescope Science Institute, 3700 San Martin Drive, Baltimore, MD 21218, USA}

\author{David Schlegel}
\affiliation{Lawrence Berkeley National Laboratory, 1 Cyclotron Road, Berkeley, CA 94720, USA}

\author{Michael Schubnell}
\affiliation{Department of Physics, University of Michigan, 450 Church Street, Ann Arbor, MI 48109, USA}
\affiliation{University of Michigan, 500 S. State Street, Ann Arbor, MI 48109, USA}

\author[0000-0002-6588-3508]{Hee-Jong Seo}
\affiliation{Department of Physics \& Astronomy, Ohio University, 139 University Terrace, Athens, OH 45701, USA}

\author[0000-0002-3461-0320]{Joseph Harry Silber}
\affiliation{Lawrence Berkeley National Laboratory, 1 Cyclotron Road, Berkeley, CA 94720, USA}

\author{David Sprayberry}
\affiliation{NSF NOIRLab, 950 N. Cherry Ave., Tucson, AZ 85719, USA}

\author[0000-0003-1704-0781]{Gregory Tarl\'{e}}
\affiliation{University of Michigan, 500 S. State Street, Ann Arbor, MI 48109, USA}

\author[0000-0003-0216-1230]{Xiaoma Wang}
\affiliation{National Astronomical Observatories, Chinese Academy of Sciences, Beijing, 100101, P.R.China}
\affiliation{School of Astronomy and Space Science, University of Chinese Academy of Sciences, Beijing, 100049, P.R.China}

\author{Benjamin Alan Weaver}
\affiliation{NSF NOIRLab, 950 N. Cherry Ave., Tucson, AZ 85719, USA}

\author{Shuo Yuan}
\affiliation{National Astronomical Observatories, Chinese Academy of Sciences, Beijing, 100101, P.R.China}

\collaboration{DESI Collaboration}

\date{\today}

\begin{abstract}
We present joint measurements of the pre- and post-reconstruction power spectra, $P_{\rm pre}$ and $P_{\rm post}$, together with their cross-power spectrum, $P_{\rm cross}$, for the Luminous Red Galaxies (LRGs) in the DESI Data Release 1 (DR1). We jointly analyse these observables with an emulator-based full-shape modeling framework, thereby, for the first time, we extract complementary nonlinear information from the galaxy density field before and after reconstruction in real survey data. Specifically, including $P_{\rm post}$ and $P_{\rm cross}$ in addition to $P_{\rm pre}$ (hereafter $P_{\rm all}$) yields an improvement of approximately $18$-$27\%$ in the $\sigma_8$ constraint in both $\Lambda$CDM and $w$CDM, depending on the redshift bin, relative to the $P_{\rm pre}$-only analysis with the cosmic microwave background distance priors (hereafter CMB). In $w$CDM, the joint CMB+$P_{\rm all}$ analysis can tighten the constraints on $w$ by approximately $5$-$15\%$ across the two LRG redshift bins, compared to the CMB+$P_{\rm pre}$ combination. Further incorporating the Type Ia supernova dataset and comparing the cosmological constraints in $w$CDM from each individual power-spectrum component with those from the full combination, we find that $P_{\rm all}$ consistently provides the tightest constraints. From the joint CMB+$P_{\rm all}$+DES-Dovekie dataset, we obtain $\Omega_m = 0.314 \pm 0.0048$ and $w = -0.988 \pm 0.023$ for the \texttt{LRG1} sample, and $\Omega_m = 0.318 \pm 0.0046$ and $w = -0.988 \pm 0.025$ for \texttt{LRG2}. These results demonstrate that combining pre- and post-reconstruction power spectra with their cross-correlation enables DESI to harvest additional nonlinear information, leading to tighter constraints on cosmological parameters.
\end{abstract}

\maketitle

\section{Introduction}
Wide-area spectroscopic galaxy surveys have become essential tools in modern cosmology, enabling precise measurements of the cosmic expansion history and the growth of large-scale structure. These key observables encode fundamental information about cosmological parameters and the nature of gravity. Over the past few decades, a series of ambitious spectroscopic surveys, like the 2dF Galaxy Redshift Survey (2dFGRS) \citep{2dFGRS}, the Sloan Digital Sky Survey (SDSS) \citep{SDSS}, the SDSS-III Baryon Oscillation Spectroscopic Survey (BOSS) \citep{BOSS}, and the SDSS-IV extended Baryon Oscillation Spectroscopic Survey (eBOSS) \citep{eBOSS}, have significantly improved the precision of large-scale structure measurements and established large-scale structure surveys as one of the cornerstones of modern observational cosmology.

Building upon these pioneering efforts, the current generation of large spectroscopic surveys is achieving unprecedented depth and coverage, pushing the limits of sampling density and surveyed cosmic volume. These include the space-based Euclid mission launched by the European Space Agency \citep{Euclid:2024yrr}, as well as the ground-based surveys like the Dark Energy Spectroscopic Instrument (DESI) \citep{DESI:2016igz,DESI:2016fyo,DESI:2022xcl,DESI:2022nlo,DESI:2023iob,DESI:2023mkx,2024AJ....168..245P} and the 4-metre Multi-Object Spectroscopic Telescope (4MOST) \citep{guiglion20194most,walcher20194most}. Together, these ongoing programs mark a major step forward in our quest to understand the cosmic expansion history and the growth of large-scale structure. In the past two years, the release of data from DESI has ushered in a new era of precision cosmology, providing the most detailed measurements to date of the large-scale structure of the Universe. In 2024, the DESI collaboration released its first-year data (called DESI DR1) \citep{DESI:2025fxa}, achieving an unprecedented 0.52\% precision in the measurement of baryon acoustic oscillations (BAO) \citep{DESI:2024uvr} based on $5.7$ million galaxy and quasar samples \citep{DESI:2024aax}. In 2025, the DESI collaboration reported a 30–50\% improvement in the statistical precision of BAO measurements relative to DESI DR1, based on its three-year dataset (called DESI DR2), with the large-scale structure sample increased to 14 million objects \citep{DESI:2025zgx}. 

While BAO measurements already provide powerful geometric constraints, a key frontier for these surveys is to robustly exploit the nonlinear information encoded in the full clustering pattern of galaxies \citep{DESI:2024jxi,DESI:2024hhd}. Obtaining reliable full-shape information on small scales remains challenging, since on large scales galaxies act as linearly biased tracers of the underlying matter distribution, whereas on smaller scales nonlinear effects make this relationship significantly more complex. Therefore, the range used for full-shape modeling is typically restricted to the mildly non-linear scale \ie\,$k =0.2\,h\,\mathrm{Mpc}^{-1}$ \citep{DESI:2024jxi,DESI:2024hhd}. Pushing beyond this regime without sacrificing robustness requires new strategies to capture nonlinear information while controlling theoretical systematics.

The BAO reconstruction technique was proposed by Eisenstein et al. \citep{Eisenstein:2006nk} to partially reverse the nonlinear evolution of the density field and thereby restore its linearity to a certain extent. This process not only improves the precision of BAO measurements but also increases the amount of cosmological information that can be extracted from post-reconstruction full-shape analyses, as the two-point statistics of the reconstructed catalogue inherently incorporate partial information from higher-order correlations\,-\,specifically, from the three- and four-point statistics of the original density field \citep{Schmittfull:2015mja}. Thus, the reconstruction technique gives surveys stronger capability to use full-shape information, not just the BAO feature. In particular, comparing and combining the pre- and post-reconstruction density fields offers a promising route to access nonlinear information that is otherwise difficult to model.

Motivated by this, Wang et al. \citep{Wang:2022nlx} proposed an approach to effectively recover nonlinear information through a joint analysis of the power spectra of the pre- and post-reconstructed density fields ($P_{\rm pre}$ and $P_{\rm post}$) and their cross-power spectrum ($P_{\rm cross}$). This method leverages the different levels of linearity between the pre- and post-reconstructed density fields, which contain complementary cosmological information and together lead to tighter parameter constraints. In practice, modeling the post-reconstructed power spectra remains challenging. Several studies have attempted to describe them within perturbation theory \citep{Hikage:2017tmm,Hikage:2019ihj,Chen:2019lpf,Sugiyama:2024eye}. Beyond perturbation theory-based approaches, our recent work has demonstrated that simulation-based numerical methods through employing an emulator framework \citep{Wang:2023hlx}, which allows one to model the reconstruction process accurately, provide an efficient and flexible avenue for extracting cosmological information within this new methodology. Taken together, these developments open a concrete pathway to harvesting nonlinear information from galaxy surveys in a controlled, survey-specific manner.

In this paper, we apply the method and emulator developed in Wang et al. \citep{Wang:2022nlx,Wang:2023hlx} to the DESI DR1 Luminous Red Galaxy (LRG) samples. Our goal is to quantify how much additional nonlinear information can be extracted by jointly analysing $P_{\rm pre}$, $P_{\rm post}$, and $P_{\rm cross}$ in a real, wide-area survey, and to assess the corresponding gains in cosmological constraints. Specifically, we present the measurements of $P_{\rm pre}$, $P_{\rm post}$, and $P_{\rm cross}$, validate our analysis pipeline using simulations, and report the cosmological constraints derived from their joint fitting. The structure of this paper is as follows. In the next section, we describe the statistical methodology. Sec.\,\ref{sec:emu} introduces the construction of the emulator. In Sec.\,\ref{sec:mockdata}, we present the mock catalogues and observational data used in this work. Sec.\,\ref{sec:result} shows the main results and cosmological constraints. Finally, Sec.\,\ref{sec:summary} provides our conclusions and discussion.

\section{Methodology}\label{sec:method}

This section outlines the statistical methodology adopted in this work and describes the procedures used to measure the galaxy power spectrum and the associated power-spectrum window functions.

\subsection{Power spectrum}

In this work, we adopt the power spectrum estimator of Ref.~\citep{Hand:2017irw}, which is based on the FKP field \citep{Feldman:1993ky}
\ba
F(\mathbf{r})=n_D(\mathbf{r})-\alpha_R n_R(\mathbf{r})\,,
\ea
where $n_D(\mathbf{r})$ and $n_R(\mathbf{r})$ are the weighted number densities of galaxies and randoms painted onto a grid, and $\alpha_R=\sum_{i=1}^{N_D} w_{d, i} / \sum_{i=1}^{N_R} w_{r, i}$ is the ratio of the total weight of galaxies to that of randoms.

The power spectrum multipoles are given by 
\ba
\begin{gathered}
\hat{P}_{\ell}(k)=\frac{2 \ell+1}{A} \int \frac{d \Omega_k}{4 \pi} F_0(\mathbf{k}) F_{\ell}(-\mathbf{k})-\mathcal{N}_{\ell}\,, \\
F_{\ell}(\mathbf{k})=\int d \mathbf{r} F(\mathbf{r}) \mathcal{L}_{\ell}(\hat{\mathbf{k}} \cdot \hat{\eta}) e^{i \mathbf{k} \cdot \mathbf{r}}\,,
\end{gathered}
\ea
where $\hat{\eta}$ denotes the line-of-sight (LOS) direction. For the cubic mocks, $\hat{\eta}$ is taken to be global, \ie, $\hat{\eta} = \hat{z}$. For the cut-sky catalogues, $\hat{\eta}$ is defined locally, \ie, $\hat{\eta} = \hat{\mathbf{r}}$, corresponding to the position vector of the first galaxy of the pair with respect to the observer. The normalization term is estimated as $A=\alpha_R \int d \mathbf{r} n_R(\mathbf{r}) n_D(\mathbf{r})$. We compute the power spectrum multipoles using the public {\tt pypower} code\footnote{\url{https://github.com/cosmodesi/pypower}}. The $k$-bin width is set to be $\Delta k = 0.01\,h\,{\rm Mpc}^{-1}$ for all the power spectrum measurements.

To perform density-field reconstruction, we adopt the {\bf RecSym} convention, \ie, both galaxies and randoms are shifted in the same way using the redshift-space displacement, thereby preserving the redshift-space distortions (RSD) in the post-reconstruction clustering. The galaxy density field is first smoothed with a Gaussian kernel $K(k)=\exp[-(k\Sigma_{\rm sm})^2/2]$ in Fourier space, adopting the optimal smoothing scale for LRGs, $\Sigma_{\rm sm}=15\,h^{-1}{\rm Mpc}$ \citep{DESI:2024uvr}. The displacement field is estimated using the Zeldovich approximation, 
\ba 
\tilde{\mathbf{s}}(\mathbf{k})=-i\frac{\mathbf{k}}{k^2}\frac{\delta({\mathbf{k}})}{b+f \mu^2}K(k)\,,
\ea
where the linear bias $b$ and growth rate $f$ used in the reconstruction algorithm are set to $b=2$ and $f=0.83$ \citep{DESI:2024uvr}. An inverse Fourier transform of $\tilde{\mathbf{s}}(\mathbf{k})$ yields the configuration-space shift field $\mathbf{s}(\mathbf{x})$, which is then applied to move both galaxies and randoms. The reconstruction is performed using {\tt pyrecon}\footnote{\url{https://github.com/cosmodesi/pyrecon}}. 

The shot noise term $N_{\ell}$ is non-zero only for the monopole. For the monopole of the auto power spectra before and after density-field reconstruction, the shot noise is subtracted as a constant term:
\ba
\mathcal{N}_0=\frac{1}{A}\left[\sum_{i=1}^{N_D} w_{D, i}^2+\alpha_R^2 \sum_{i=1}^{N_R} w_{R, i}^2\right]\,.
\ea
The shot noise in the cross-power spectrum, $P_{\rm cross}$, is scale-dependent \citep{Wang:2022nlx}. It is estimated using the ``half-sum/half-difference'' (HS/HD) approach and subsequently subtracted from the measured spectrum \citep{Ando:2017wff}.

\subsection{Power spectrum window}

When comparing power spectrum measurements from galaxy surveys with theoretical (or emulator-based) predictions, it is essential to accurately model the survey footprint and selection function to avoid biased cosmological constraints. This is typically achieved using a window matrix that encodes the survey geometry and relates the measured quantities, $\hat{P}_{\ell}(k)$, to the theoretical (or emulator-based) predictions, $P_{\ell'}(k')$, as
\begin{align}
\hat{P}_{\ell}(k)=\sum_{\ell'}\sum_{k'}W_{\ell \ell'}(k, k')P_{\ell'}(k'),
\end{align}
where the window matrix $W_{\ell \ell'}(k, k')$ can be computed from random catalogs following \citep{Beutler:2021eqq,Pinon:2024wzd}. It is given by
\begin{widetext}
\begin{align}
&W_{\ell \ell^{\prime}}^{(n)}\left(k, k^{\prime}\right)=\frac{i^{\ell}(-i)^{\ell^{\prime}}}{2 \pi^2}
\int s^2d s \sum_p \frac{2 \ell+1}{2 p+1} A_{p \ell \ell^{\prime}}
W_p^{(n)}(s) j_{\ell^{\prime}}\left(k^{\prime} s\right) j_{\ell}(k s),\
&\mathcal{L}_{\ell}(\mu)\mathcal{L}_{\ell^{\prime}}(\mu)=\sum_p A_{p \ell \ell^{\prime}} \mathcal{L}_p(\mu),
\end{align}
\end{widetext}
where $W_p^{(n)}(s)$ denotes the configuration-space window function, and the superscript $(n)$ indicates that the window matrix includes the $n$th-order wide-angle (WA) corrections. The WA effect arises when the pair separation of galaxies becomes non-negligible compared to their distance from the observer, breaking the plane-parallel (or distant-observer) approximation and introducing corrections that depend on the varying LOS direction across the sky. In this analysis, we include terms up to first order in the WA expansion. 

Furthermore, by using the relation between the first-order WA power spectrum and its zeroth-order counterpart, $P_{\ell}^{(1)}(k)=\sum_{\ell^{\prime}} W_{\ell \ell^{\prime}}^{\mathrm{WA},(1)}(k) P_{\ell^{\prime}}^{(0)}(k)$, the WA correction can be incorporated into the total window function as \citep{Pinon:2024wzd},
\begin{widetext}
\ba
\begin{gathered}
W_{\ell \ell^{\prime}}\left(k, k^{\prime}\right)=W_{\ell \ell^{\prime}}^{(0)}\left(k, k^{\prime}\right)+\frac{1}{k^{\prime}} \sum_{\ell^{\prime \prime}} W_{\ell \ell^{\prime \prime}}^{(1)}\left(k, k^{\prime}\right) W_{\ell^{\prime \prime} \ell^{\prime}}^{\mathrm{WA},(1)}\left(k^{\prime}\right)\\
W_{\ell \ell^{\prime}}^{\mathrm{WA},(1)}(k)=-i \frac{\ell(\ell-1)}{2(2 \ell-1)}\left[(\ell-1)-k \partial_k\right] \delta_{\ell-1, \ell^{\prime}}-i \frac{(\ell+1)(\ell+2)}{2(2 \ell+3)}\left[(\ell+2)+k \partial_k\right] \delta_{\ell+1, \ell^{\prime}}.
\end{gathered}
\ea
\end{widetext}
We compute the power-spectrum window matrices using {\tt pypower}.

\section{Building the emulator} \label{sec:emu}

We follow the methodology developed in Wang et al.~\citep{Wang:2022nlx} and construct our emulator using the \textsc{Dark Quest} simulation suite \citep{Nishimichi:2018etk}. The simulations consist of $N$-body runs with $2048^3$ dark matter particles in a $2\,h^{-1}\mathrm{Gpc}$ box. Dark matter halos are identified using the phase-space temporal friends-of-friends halo finder, \textsc{Rockstar} \citep{behroozi2012rockstar}. We use simulation snapshots at redshifts $z=0.484$ and $z=0.689$, chosen to closely match the effective redshifts of the first two DESI LRG bins. The \textsc{Dark Quest} suite covers $100$ spatially flat $w$CDM cosmologies with six varying parameters, as well as one $\Lambda$CDM realization consistent with the \textit{Planck} 2015 best-fit parameters \citep{Planck:2015fie}.

We populate galaxies into dark matter halos using the five-parameter Halo Occupation Distribution (HOD) prescription of Zheng et al., \citep{Zheng:2007zg}. As in Wang et al.~\citep{Wang:2022nlx}, we reparametrize the HOD model in terms of four free parameters, $\{\log_{10}\sigma_{\log M}, \log_{10}(M_0/M_1), \alpha, M_1/M_{\min}\}$. The meanings and sampling ranges of both the cosmological and HOD parameters are listed in Table~\ref{tab:para}. For each parameter set, $M_{\min}$ is computed by solving for a fixed galaxy number density of $n = 3.6\times10^{-4}\,(h/{\rm Mpc})^{3}$, ensuring that the simulated catalogues match the number density of the DESI LRGs and yielding a consistent five-parameter HOD configuration for every mock sample.

\begin{table*}[ht!]
\centering
\renewcommand{\arraystretch}{1.3}
\resizebox{\textwidth}{!}{%
\begin{tabular}{ccc}
\hline\hline
Parameter  & Meaning &Sampling range     \\\hline
$\omega_b \equiv \Omega_b h^2 $   &the physical energy density of baryon& $[0.0211375,0.0233625]$\\
$\omega_c \equiv \Omega_c h^2$   &the physical energy density of cold dark matter&$[0.10782,0.13178]$\\
$\Omega_{de}$  &the dimensionless energy density of dark energy&$[0.54752,0.82128]$\\
$\ln(10^{10}A_s)$   & the amplitude of the primordial power spectrum&$[2.4752,3.7128]$\\
$n_s$  & the spectral index of the primordial power spectrum&$[0.916275,1.012725]$\\
$w$   &the equation of state of dark energy&$[-1.2,-0.8]$\\\hline
$\log \sigma_{{\rm log}M}$  & logarithmic steepness of the central-galaxy occupation transition& $[-2.99,0.48]$ \\
$\log (M_0/M_1)$  &logarithmic ratio of the cutoff mass $M_0$ to the satellite normalization mass $M_1$& $[-2.99,-0.4]$ \\
$\alpha$  &the power-law index on the number of satellites & $[0.2,1.5]$ \\
$M_1/M_{{\rm min}}$  &ratio between $M_1$ and the minimum mass for hosting a central galaxy $M_{\rm min}$ & $[6,18]$ \\ 
\hline\hline
\end{tabular}
}
\renewcommand{\arraystretch}{1}
\caption{Meanings and sampling ranges of the cosmological (upper block) and HOD (lower block) parameters that define the parameter space used to construct the emulator.}
\label{tab:para}
\end{table*}

A total of $2450$ HOD realizations are sampled within the prior ranges using a Sobol low-discrepancy sequence \citep{Sobol:1967}. Among them, $2400$ realizations are assigned to $80$ cosmologies for training, \ie, each training cosmology is associated with $30$ distinct HODs. The remaining $50$ realizations are assigned to each testing cosmology, yielding a testing set of $1000$ models. When measuring the power spectra from training and testing sets (as described in Sec.\,\ref{sec:method}), we account for the Alcock-Paczynski (AP) effect \citep{AP} by rescaling the simulation coordinates according to AP factors ($ q_{\perp}, q_{\|}$) \citep{Wang:2022nlx}, which quantify the discrepancy between the true transverse and line-of-sight distances and those inferred under the fiducial cosmology. 

With the measured power-spectrum training set, we construct the emulator using a Gaussian Process (GP). The GP training involves selecting a Mat\'ern kernel to model the correlations among training points and optimizing the kernel hyperparameters by maximizing the log-likelihood (see Eq.~8 of Wang et al.~\citep{Wang:2022nlx}). Once trained, the GP can predict the power spectrum at new parameter points.

The $1000$ testing realizations are used to assess the emulator prediction error across the parameter space. The resulting emulator uncertainty is propagated into the cosmological inference by incorporating it into the likelihood, in combination with the statistical covariance of the power-spectrum measurements.

\section{Mocks and data} \label{sec:mockdata}

In this section, we describe the mock catalogs and observational data used in our analysis. We first summarize the mock samples employed for emulator validation, systematic-error assessment, and covariance estimation, and then present the DESI LRG data and the external datasets used for cosmological inference.

\vspace{0.6em}
\mockitem{\textsc{Dark Quest}}{
We use $15$ realizations of the \textit{Planck} 2015 $\Lambda$CDM cosmology from the \textsc{Dark Quest} simulation suite \citep{Nishimichi:2018etk} as one of our validation mock sets. The simulation specifications and the halo-galaxy population model, which are identical to those adopted for building the emulator, have been described in Sec.\,\ref{sec:emu}. In this section, we use the \textsc{Dark Quest} fiducial mocks at redshifts $z = 0.484$, and $ 0.689$ to assess the accuracy of the emulator predictions and to test the robustness of the full-shape fitting pipeline.
}

\vspace{0.6em}
\mockitem{\textsc{AbacusSummit}}{
The \textsc{AbacusSummit} suite consists of high-resolution $N$-body simulations performed with the \textsc{Abacus} code \citep{Maksimova:2021ynf}. Each simulation evolves $6912^3$ dark matter particles in a $(2\,h^{-1}{\rm Gpc})^3$ periodic box, assuming the \textit{Planck} 2018 $\Lambda$CDM cosmology. Dark matter halos are identified using the \textsc{CompaSO} halo finder \citep{Hadzhiyska:2021zbd}, and galaxies are populated following the LRG HOD model \citep{Yuan:2023ezi}. A second generation of the Abacus mocks was produced (referred to as {\tt Abacus-2}) whose HOD have been fitted to the DESI Early Data Release (EDR) measurements at redshifts $z = 0.5$, $0.8$, and $1.1$ \citep{Yuan:2023ezi}. The latest mock set, called {\tt Abacus-High Fidelity (HF)} , is constructed with HOD parameters calibrated to the DESI DR2 data at redshifts $z = 0.5$, $0.725$, and $0.9$. This study employs $25$ realizations of the \textsc{AbacusSummit} simulations for mock testing and systematic-error assessment. Three sets of galaxy mocks are used:
\begin{itemize}
\item \textbf{{\tt Abacus-2 complete}:} galaxy mocks generated without fibre-assignment incompleteness, available in both cut-sky and cubic formats. Together, these two versions are used to test the robustness and accuracy of the power-spectrum window-matrix calculation and to assess potential biases arising from survey geometry;
\item \textbf{{\tt Abacus-2 altMTL}:} galaxy mocks generated using the same {\tt fibreassign} algorithm as applied to the real DESI observations, thereby faithfully reproducing the fibre-assignment effects present in the DESI DR1 data. Together with the {\tt Abacus-2 complete} cut-sky mocks, they enable a quantitative assessment of the impact of fibre-assignment incompleteness on the measured power spectra;
\item \textbf{{\tt Abacus-HF}:} compared to the {\tt Abacus-2} mocks, the second and third redshift bins of the {\tt Abacus-HF} mock set are closer to the effective redshifts of the DESI LRG observations. In this work, we use only the {\tt Abacus-HF} cut-sky mocks.
\end{itemize}
In summary, these {\tt Abacus} mocks are primarily used to validate the full-shape fitting pipeline, and to quantify observational systematics arising from fibre-assignment incompleteness, in a setup that closely mirrors the DESI survey.
}

\vspace{0.6em}
\mockitem{\textsc{EZmock}}{
The \textsc{EZmock} catalogs \citep{chuang2015ezmocks} are fast approximate mocks generated using a semi-empirical structure formation algorithm calibrated on $N$-body simulations. Each realization reproduces the DESI DR1 survey geometry, angular mask, and radial selection function of the LRG sample. In this analysis, we use two sets of \textsc{EZmock}:
\begin{itemize}
\item \textbf{{\tt EZmock${1000}$}:} a suite of $1000$ realizations employed to estimate the covariance matrix of the pre-, post-, and cross-power spectra. Both the Hartlap and Percival correction factors \citep{Hartlap:2006kj, Percival:2021cuq} are applied to debias the inverse covariance and to account for the additional uncertainty arising from the finite number of mocks, thereby preventing an underestimation of the parameter errors;
\item \textbf{{\tt EZmock${50}$}:} a set of 50 realizations generated both with and without the radial integral constraint (RIC) \citep{deMattia:2019vdg}. The RIC arises because the redshift distribution of random catalogs is constructed to exactly match that of the data, which effectively removes (or “nulls”) radial modes in the measured power spectra. By comparing the RIC and no-RIC versions, this mock set enables a direct quantification of the RIC impact on the DESI LRG power-spectrum measurements.
\end{itemize}
}

\vspace{0.6em}
\mockitem{\textsc{DESI DR1 LRGs}}{
The DESI DR1 Luminous Red Galaxy sample used in this analysis consists of approximately $2.14$ million reliable redshifts over $5{,}840\,{\rm deg}^2$ within the redshift range $0.4 < z < 1.1$ \citep{DESI:2024uvr}. For clustering analysis, the sample is divided into three disjoint redshift bins: $0.4<z<0.6$ (\texttt{LRG1}), $0.6<z<0.8$ (\texttt{LRG2}), and $0.8<z<1.1$ (\texttt{LRG3}). Across the first two redshift bins, the LRG selection yields a nearly constant comoving number density of $n \simeq 3.6 \times 10^{-4}\,h^{3}\mathrm{Mpc}^{-3}$ as shown in Fig.\,\ref{fig:1}. The highest-redshift bin overlaps with the DESI emission line galaxy (ELG) sample, offering prospects for future joint multi-tracer analyses. In our analysis, given the significant variation in number density within the highest-redshift bin, we restrict our analysis in the emulator framework to the first two bins, \texttt{LRG1} and \texttt{LRG2}. The effective redshifts of these two bins are $z_{\mathrm{eff}} = 0.51$ and $0.71$ respectively.

\begin{figure}[htp]
    \centering
   \includegraphics[width=0.45\textwidth]{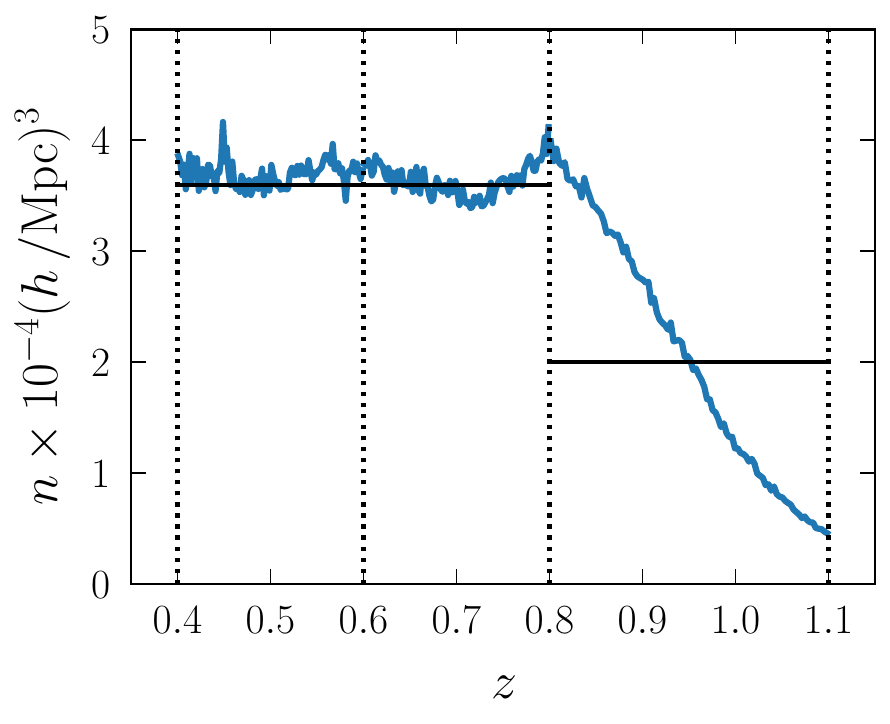}
    \caption{Comoving number density as a function of redshift for the DESI DR1 LRG sample. The horizontal lines show the mean number densities in three redshift bins, illustrating the nearly constant number density in the first two bins that motivates our emulator setup.}
    \label{fig:1}
\end{figure}

Various observational systematics are corrected by incorporating appropriate weights into the data catalogues \citep{DESI:2024aax}. Each galaxy is assigned a total weight ($w_{\rm tot}$) \citep{DESI:2024aax}, including imaging systematics, targeting completeness, and variations in the instrument’s redshift success rate, multiplied by the FKP weight ($w_{\rm FKP}$) \citep{Feldman:1993ky}. We also apply a set of mock-based corrections to mitigate residual observational effects. First, the fibre assignment (FA) incompleteness is corrected \footnote{For the pre-reconstruction sample, an alternative correction based on a $\theta$-cut has been applied \citep{Pinon:2024wzd}. Extending the $\theta$-cut scheme to the post-reconstruction sample remains a topic for future investigation. In this work, we therefore adopt a simulation-based correction method for both the pre- and post-reconstruction samples to ensure a consistent treatment.} using the {\tt Abacus-2} mocks through
\begin{equation}
    \Delta P_{\mathrm{FA}} \equiv P_{\mathrm{altMTL}}^{\mathrm{mock}} - P_{\mathrm{complete}}^{\mathrm{mock}},
\end{equation}
where the difference between the {\tt altMTL} and {\tt complete} mocks quantifies the fibre-assignment effect. Second, the correction to RIC effect is estimated using the {\tt EZmock50} realizations,
\begin{equation}
    \Delta P_{\mathrm{RIC}} \equiv P_{\mathrm{RIC}}^{\mathrm{mock}} - P_{\mathrm{no\text{-}RIC}}^{\mathrm{mock}}.
\end{equation}
Finally, we assess the impact of the angular integral constraint (AIC) introduced by the imaging systematic weights; this effect is found to be negligible for the LRG sample \citep{DESI:2024aax} and is therefore not corrected for in our baseline analysis.
}

\vspace{0.6em}
\mockitem{\textsc{External Data}}{In addition to the DESI LRG data, we include two external datasets to improve cosmological parameter constraints in this analysis:
\begin{itemize}
\item \textbf{CMB distance priors:} We adopt the compressed CMB distance information derived from the \textit{Planck} 2018 temperature and polarization data \citep{Chen:2018dbv}. The four parameters, \ie\, the shift parameter $R$, the acoustic scale $l_A$, the physical baryon density $\Omega_b h^2$, and the scalar spectral index $n_s$, capture the key geometric and early-Universe information of the CMB, providing an efficient summary of the \textit{Planck} likelihood that is well suited for combination with late-time large-scale structure data.
\item \textbf{Type Ia supernovae:} We include three complementary SNe~Ia datasets to constrain the late-time expansion history:
    \begin{itemize}
        \item PantheonPlus \citep{Brout:2022vxf}: 1,550 spectroscopically confirmed SNe~Ia in the redshift range $0.001<z<2.26$, incorporating full statistical and systematic covariance;   
        \item Union3 \citep{Rubin:2023jdq}: 2,087 SNe~Ia, many overlapping with PantheonPlus, with a re-analysis of systematic uncertainties using a Bayesian hierarchical model;
        \item DES-Dovekie \citep{DES:2025sig}: a fully recalibrated SNe Ia sample containing 1623 SNe~Ia from the Dark Energy Survey Year~5 sample and 197 low-$z$ SNe~Ia from other surveys, providing an improved dataset for cosmological constraints.
    \end{itemize}
\end{itemize}
}

\begin{figure*}[htp]
    \centering
    \includegraphics[width=0.9\textwidth]{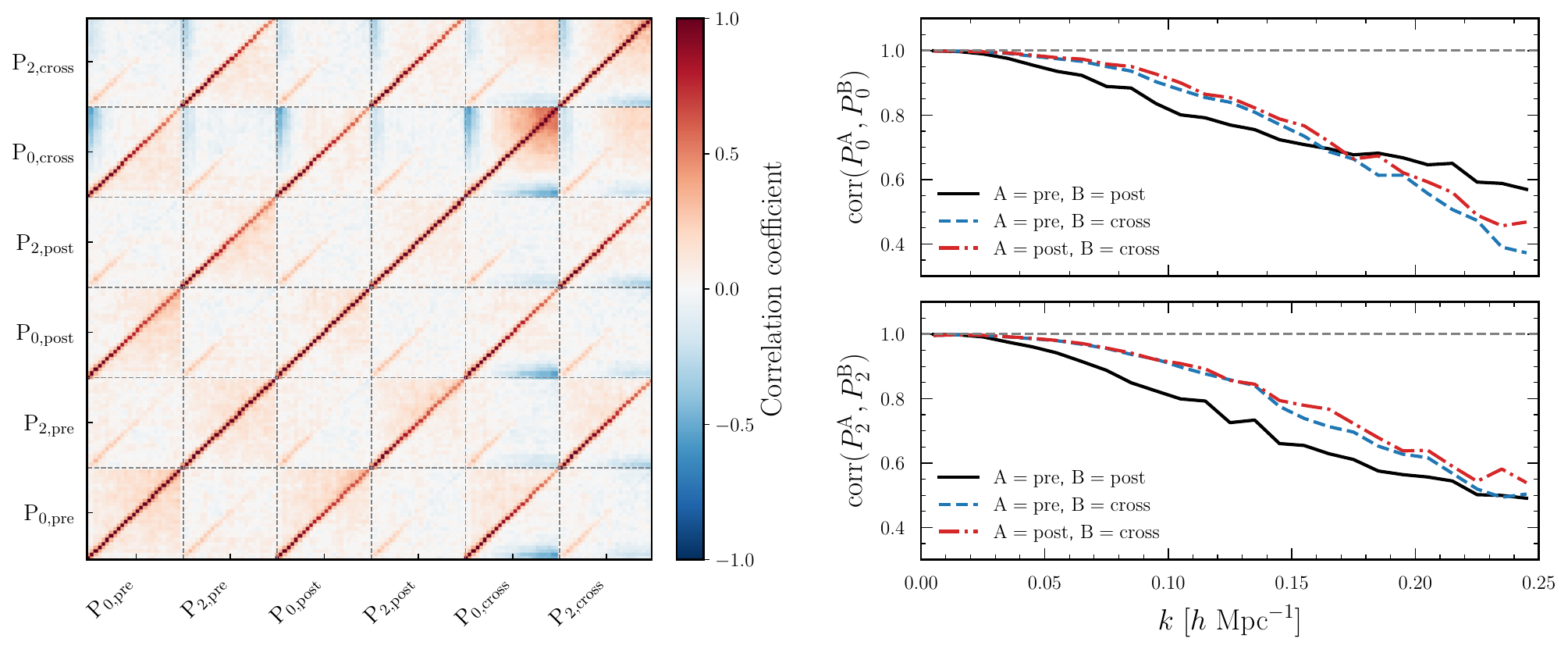}
    \caption{Left panel: Correlation matrix of the pre-, post-, and cross-reconstruction power-spectrum multipoles for the DESI DR1 LRG sample. Each block corresponds to 25 $k$-bins of $P_0$ or $P_2$ from the pre-, post-, or cross-reconstructed catalogs. The color scale indicates the correlation coefficient between multipole bins, highlighting the strong correlations between different $k$-modes and between the three types of spectra. Right panels: Correlation coefficients between the monopole (upper) and quadrupole (lower) from various types of power spectra: pre vs. post (solid black), pre vs. cross (blue dashed), and post vs. cross (red dot-dashed). The dashed horizontal line shows a perfect correlation (i.e., corr = 1) for a reference.}
    \label{fig:2z1}
\end{figure*}
\begin{figure*}[htp]
    \centering
    \includegraphics[width=0.9\textwidth]{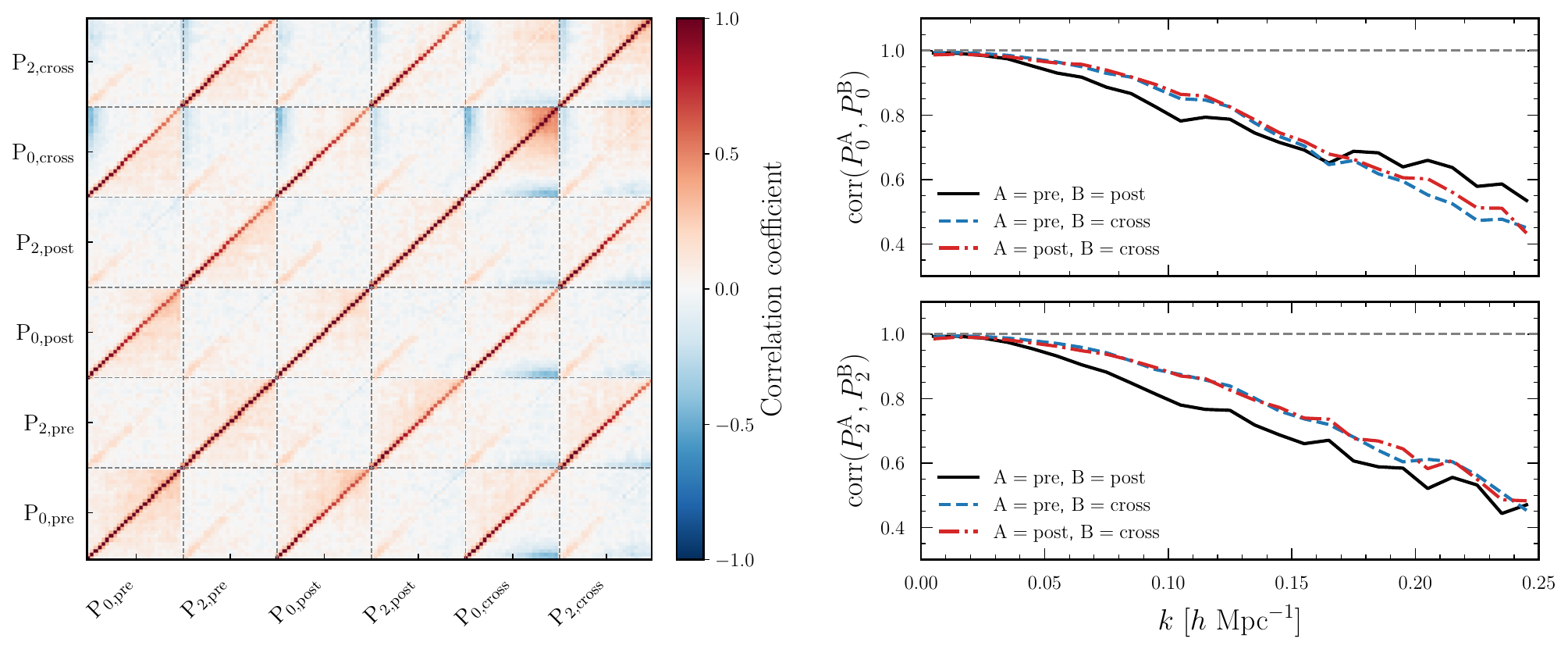}
    \caption{Same as Fig.~\ref{fig:2z1}, but for \texttt{LRG2}.}
    \label{fig:2z2}
\end{figure*}

\section{Results} \label{sec:result}

In this section, we present the main results of our analysis. We first validate the full-shape fitting pipeline using cubic mock catalogs, which provide a controlled environment free from survey geometry and observational effects. We then test the accuracy of the power-spectrum window convolution and further validate the pipeline using cut-sky mock catalogs, ensuring that the window matrix correctly accounts for the survey geometry. Finally, we apply the validated pipeline to the DESI DR1 LRG data to derive cosmological constraints.

\subsection{Cubic mock tests}

To perform the cosmological inference, we run Markov Chain Monte Carlo (MCMC) analyses using \textsc{Cobaya} \citep{Torrado:2020dgo}. For a given set of model parameters $\boldsymbol{\theta}$, the likelihood is assumed to be Gaussian, with the $\chi^2$ defined as
\begin{align}
\chi^{2}(\boldsymbol{\theta}) =
\big[ \mathbf{P}_{\rm d} - \mathbf{P}_{\rm e}(\boldsymbol{\theta}) \big]^{\top}(\mathbf{C}+\sigma^2\mathbf{I})^{-1} \big[ \mathbf{P}_{\rm d} - \mathbf{P}_{\rm e}(\boldsymbol{\theta}) \big],
\label{eq:chi2}
\end{align}
where $\mathbf{P}_{\mathrm{d}}$ denotes the measured pre-, post-, and cross-power spectra, and $\mathbf{P}_{\mathrm{e}}(\boldsymbol{\theta})$ is the emulator prediction. 

The covariance matrix $\mathbf{C}$ is computed from the {\tt EZmock${1000}$} realizations. The normalized correlation matrix is presented in left panels of Figs.~\ref{fig:2z1}-\ref{fig:2z2}. Meanwhile, the right panels of Figs.~\ref{fig:2z1}-\ref{fig:2z2} display the correlation coefficient between various types of power spectrum multipoles. The correlations between the pre-reconstructed, post-reconstructed, and cross power spectra decrease systematically with increasing wavenumber $k$. While the correlation is close to unity on the largest, linear scales, it progressively weakens toward smaller scales, reflecting the increasing impact of nonlinear evolution and reconstruction-induced mode decorrelation. This scale-dependent loss of correlation highlights the complementarity of $P_{\rm pre}$, $P_{\rm post}$, and $P_{\rm cross}$, thereby motivating their joint use to enable access to complementary nonlinear information beyond a single statistic. $\sigma$ represents the emulator prediction uncertainty estimated from the $1000$ testing realizations. This emulator uncertainty is added as an extra diagonal term in the covariance matrix, ensuring that it is consistently propagated into the cosmological constraints.

Fig.~\ref{fig:A1} and Fig.~\ref{fig:A2} in the Appendix present the residuals between the emulated and measured power spectrum multipoles, normalized by the statistical uncertainty. Fig.~\ref{fig:A3} and Fig.~\ref{fig:A4} in the Appendix show the statistical, emulation, and total uncertainties as functions of $k$.  It's found that the emulator maintains accuracy at the level of the $1\,\sigma$ statistical uncertainty over a wide scale range, depending the type of power spectra. For the $P_{\rm pre}$ quadrupole, the emulator accuracy gradually exceeds the $1\,\sigma$ statistical uncertainty at scales beyond $k \sim 0.3\,h\,\mathrm{Mpc}^{-1}$, where nonlinear effects become increasingly significant. In comparison, the emulator achieves higher predictive precision for $P_{\rm post}$ and $P_{\rm cross}$, reflecting their reduced nonlinear complexity and improved modeling accuracy after reconstruction. 

The parameter set sampled in the MCMC includes the cosmological parameters and HOD parameters, as summarized in Table~\ref{tab:para}. In the absence of external datasets, we impose additional Gaussian priors on $\Omega_{\mathrm{b}}h^{2}$ and $n_{\rm s}$ based on the \textit{Planck} 2018 measurements, since these parameters cannot be tightly constrained by the full-shape information alone.

We validate the full-shape pipeline using the cubic {\tt Dark Quest} mocks at two redshift outputs, $z=0.484$ and $z=0.689$, which match the redshifts used in our emulator construction, as well as the cubic {\tt Abacus-HF} mocks. Because these mocks are free from survey geometry and observational systematics, they provide an ideal test of whether the pipeline can recover the input cosmology under controlled conditions.

We test different $k_{\rm max}$ values, using the joint power spectra of pre+post+cross from the {\tt Dark Quest} mocks and {\tt Abacus-HF} mocks to constrain the $\Lambda$CDM model. The parameter results for $k_{\rm max}=0.2, 0.25, 0.3\,h\,\mathrm{Mpc}^{-1}$ are shown in Fig.\,\ref{fig:A11} in the Appendix. Overall, no significant systematic shift is observed as $k_{\rm max}$ increases. We note, however, that at $k_{\rm max}=0.3 \,h\,\mathrm{Mpc}^{-1}$ the constraint on the $\Omega_{de}$ parameter from the {\tt Abacus-HF} mocks at $z=0.725$ lies at the edge of the $1\,\sigma$ interval, while the remaining parameters remain well within $1\,\sigma$. Therefore, we adopt a conservative choice of $k_{\rm max}= 0.25\,h\,\mathrm{Mpc}^{-1}$ for the cosmological analysis.

\begin{figure*}[htp]
    \includegraphics[width=0.9\textwidth]{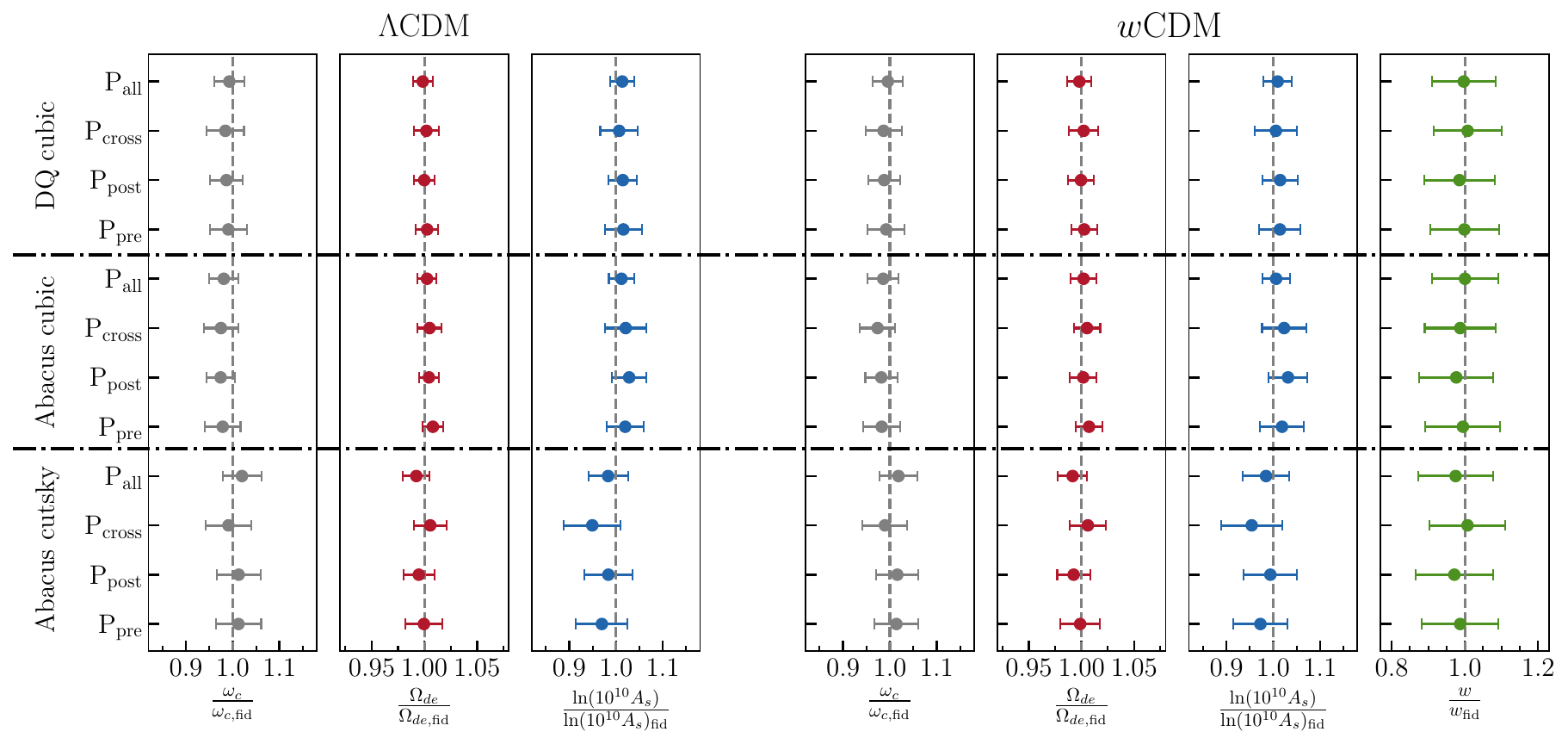}
    \caption{Mock-test results for the first redshift bin under different modeling setups. Left panels show results in the $\Lambda$CDM model, while right panels correspond to the $w$CDM model. From top to bottom, the three blocks (separated by dot-dashed lines) correspond to Dark Quest (DQ) cubic mocks, Abacus cubic mocks, and Abacus cut-sky mocks, respectively. Within each block, constraints are shown for analyses using $P_{\rm pre}$, $P_{\rm post}$, $P_{\rm cross}$, and $P_{\rm all}$. Points and error bars denote the posterior mean values and $1\sigma$ uncertainties, normalised by the fiducial parameter values, while the vertical dashed lines indicate the fiducial cosmology of the mocks. The overall consistency across different mock sets and power-spectrum combinations demonstrate the robustness of the full-shape fitting pipeline.}
    \label{fig:mockz1}
\end{figure*}

\begin{figure*}[htp]
    \includegraphics[width=0.9\textwidth]{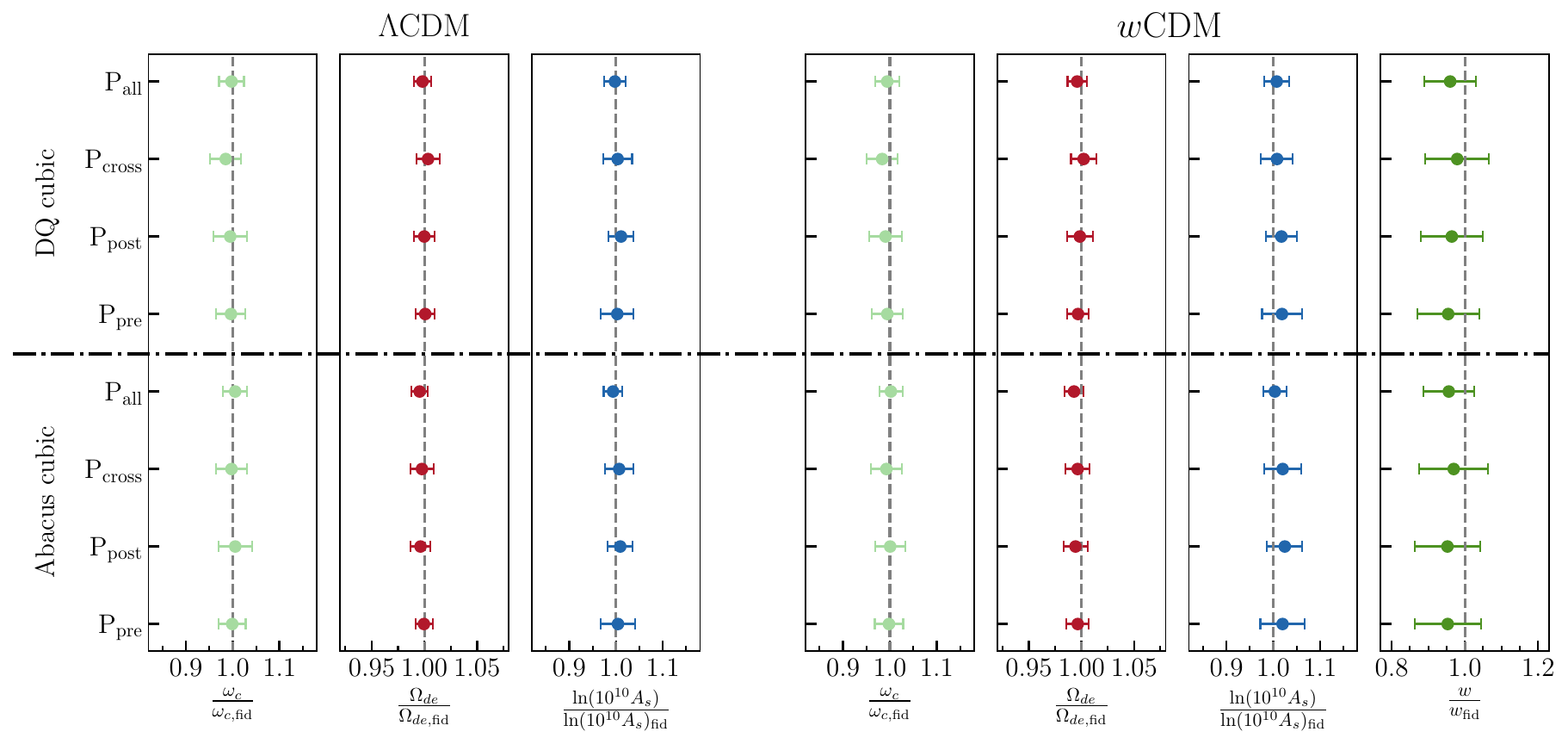}
    \caption{Same as Fig.\,\ref{fig:mockz1}, but for the second redshift bin. Since the redshift ($z=0.8$) of the second $z$ bin of {\tt Abacus-2} cutsky mocks differs substantially from that of the emulator ($z=0.689$), we do not perform a mock validation for that bin in this work.}
    \label{fig:mockz2}
\end{figure*}

First, we present the {\tt Dark Quest} mock test results from different type of power spectra. As shown in the upper-left blocks of Figs.\,\ref{fig:mockz1} and \ref{fig:mockz2}, the fiducial $\Lambda$CDM parameters are accurately reproduced. The corresponding contour plots are presented in Fig.\,\ref{fig:A5} in the Appendix. When combining the pre-reconstructed, post-reconstructed, and their cross power spectra, we obtain the tightest cosmological constraints, demonstrating the complementarity of the three statistics and the additional nonlinear information captured by their joint analysis.

To further assess the robustness of the method, we repeat the analysis while allowing the equation-of-state (EoS) parameter of dark energy, $w$, to vary. The resulting constraints, presented in the upper-right blocks of Figs.\,\ref{fig:mockz1} and \ref{fig:mockz2}, along with the contour plots in Fig.~\ref{fig:A6} in the Appendix, also successfully recover the underlying $w$CDM cosmology. These tests confirm that the pipeline reliably retrieves the input cosmology even when the additional degree of freedom $w$ is introduced.

Next, we test the robustness of the pipeline using the average power spectra of the $25$ cubic {\tt Abacus-HF} mock realizations. These simulations have redshifts that are closer to those of the DESI LRG sample, although they do not exactly coincide with the redshifts covered by the emulator. The constraining results shown in the middle-left blocks of Figs.\,\ref{fig:mockz1} and \ref{fig:mockz2}, as well as in Fig.~\ref{fig:A7} in the Appendix demonstrate that the fiducial $\Lambda$CDM cosmology is accurately reproduced. We further repeat the analysis allowing the dark-energy EoS parameter $w$ to vary. As illustrated in the middle-right blocks of Figs.\,\ref{fig:mockz1} and \ref{fig:mockz2}, as well as in Fig.~\ref{fig:A8} in the Appendix, the resulting constraints successfully recover the input $w$CDM cosmology. Together, the combination of pre-, post-, and cross-reconstructed power spectra yields the most stringent constraints on both $\Lambda$CDM and $w$CDM parameters. These mock tests indicate that the modest redshift difference between the emulator and the {\tt Abacus-HF} mocks does not lead to any noticeable bias at the current level of statistical precision.

\subsection{Validation of window matrix and cut-sky mock tests}

\begin{figure*}[htp]
    \centering
    \includegraphics[width=0.45\textwidth]{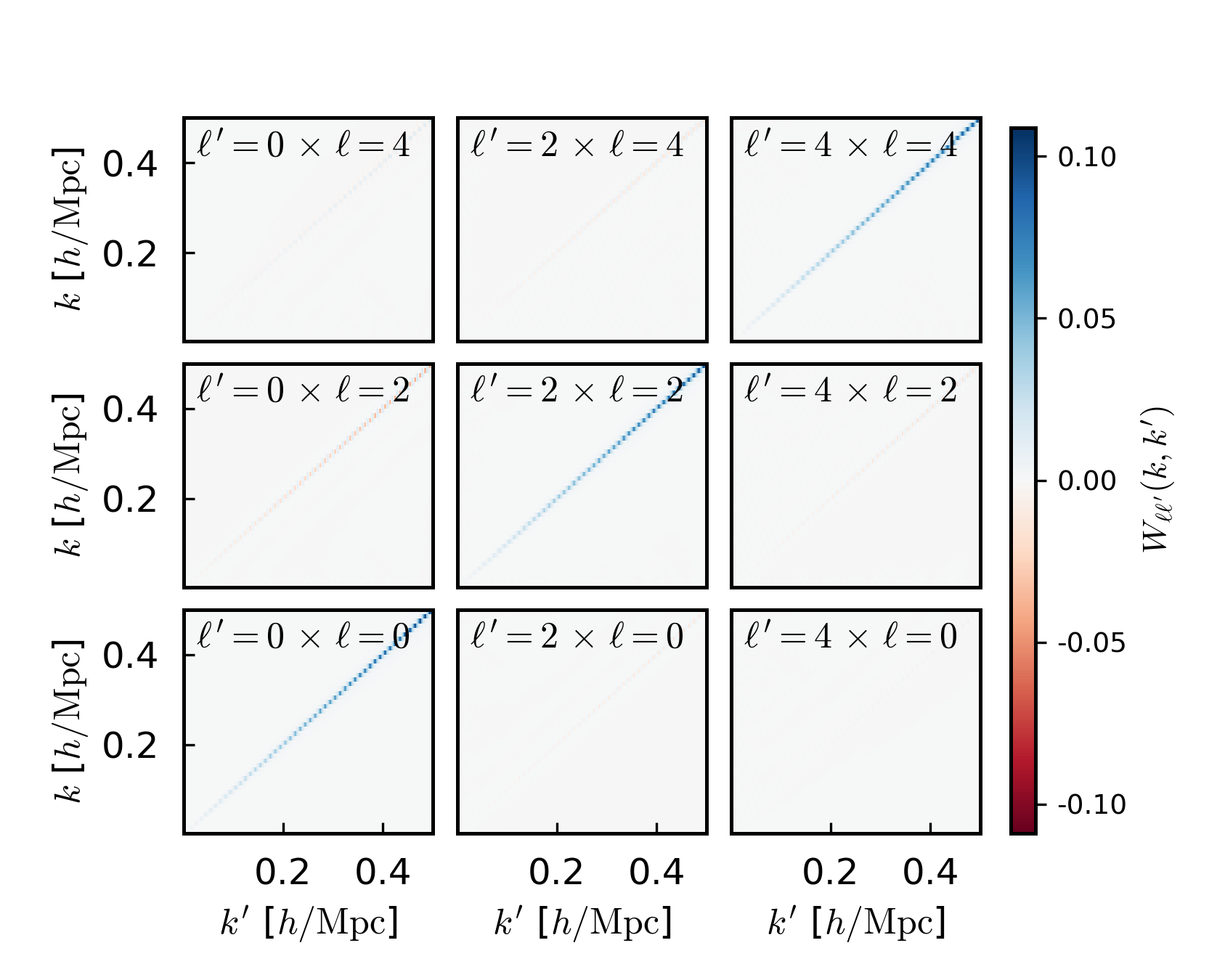}
    \includegraphics[width=0.45\textwidth]{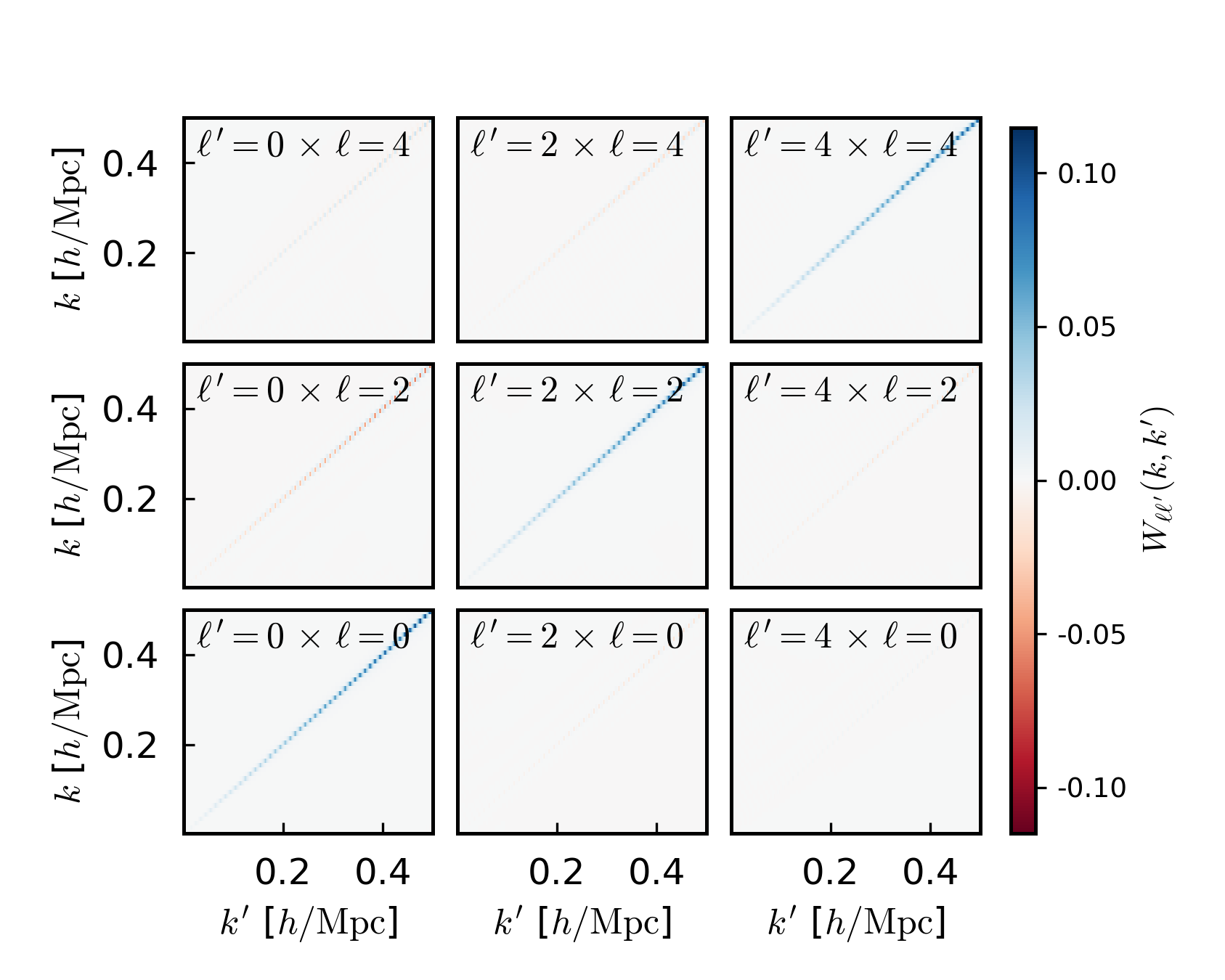}
    \caption{Window matrix for the DESI DR1 LRG survey geometry, shown for $k, k' < 0.5\,h\,\mathrm{Mpc}^{-1}$ at the effective redshifts of \texttt{LRG1} (left) and \texttt{LRG2} (right). The color scale encodes the coupling between different $(k,\ell)$ modes induced by the survey footprint and selection function.}
    \label{fig:7}
\end{figure*}

As described in Sec.~\ref{sec:method}, we compute the window matrix corresponding to the DESI DR1 survey geometry, with its structure illustrated in Fig.~\ref{fig:7}. We now validate the accuracy of this window-function implementation and its impact on the recovered cosmological parameters.

Our first test uses the average power spectrum of the 25 cubic {\tt Abacus-2 complete} mock realizations. Because these mocks contain no survey geometry, their cubic power-spectrum multipoles over the range $0.001 < k' < 0.5\,h \mathrm{Mpc}^{-1}$ can be regarded as the theoretical prediction. We therefore convolve these cubic power-spectrum multipoles with the computed window matrix to generate the expected power-spectrum multipoles for the corresponding cut-sky {\tt Abacus-2 complete} mocks. 

\begin{figure*}[htp]
    \centering
    \includegraphics[width=0.9\textwidth]{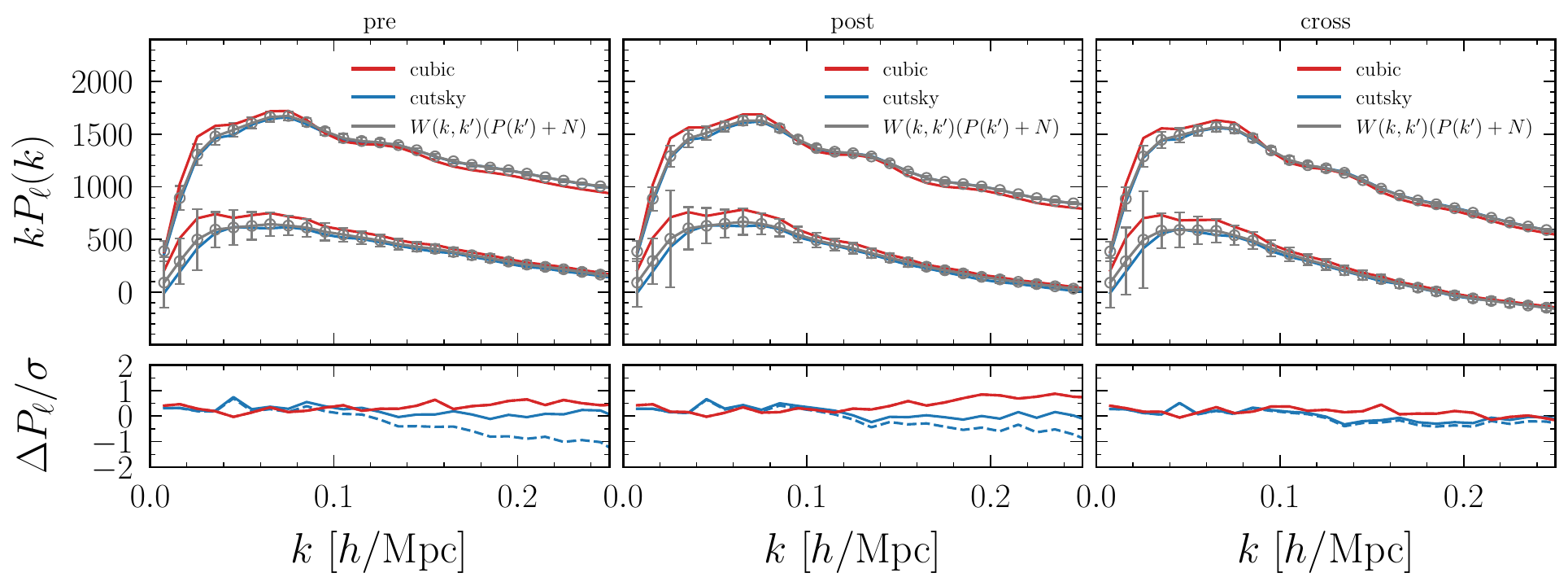}
    \caption{Top: average power-spectrum multipoles of the \textsc{AbacusSummit} cubic mocks (red lines) and cut-sky mocks (blue lines) at $z=0.5$. Gray lines show the convolved power-spectrum multipoles obtained by applying the window matrix to the cubic-mock measurements and adding a constant monopole term. Bottom: difference between the measured monopole (blue lines) and quadrupole (red lines) from cut-sky mocks and the convolved prediction, normalized by the standard deviation of the DESI DR1 power spectrum. The solid (resp. dashed) lines are the convolved predictions with (resp. without) a constant monopole term added.}
    \label{fig:8}
\end{figure*}

\begin{figure*}[htp]
    \centering
    \includegraphics[width=0.9\textwidth]{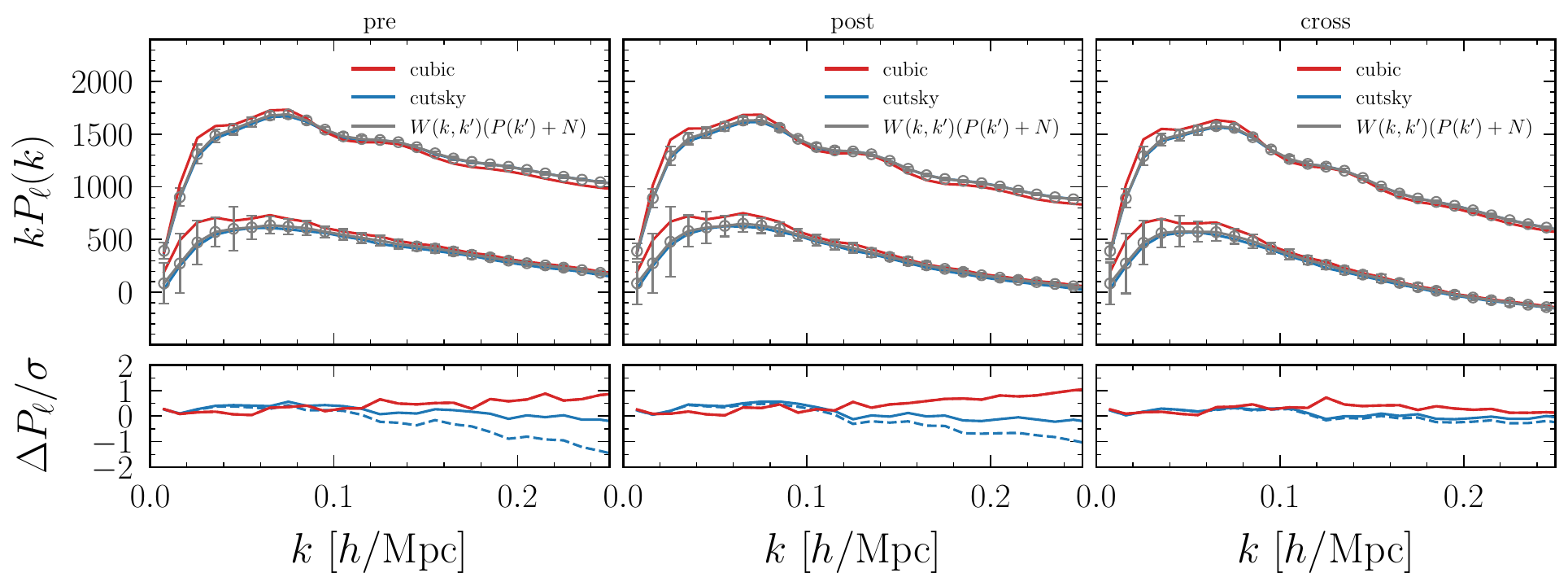}
    \caption{Same as Fig.\,\ref{fig:8}, but at $z=0.8$.}
    \label{fig:9}
\end{figure*}

The dashed lines in the lower panels of Figs.~\ref{fig:8} and \ref{fig:9} show the difference between the cubic power-spectrum multipoles multiplied by the window matrix and the cut-sky power-spectrum multipoles, normalized by the standard deviation. The residuals increase in amplitude toward smaller scales, indicating a larger discrepancy at high $k$.

To further reduce the residuals at small scales, we introduce a free constant term $N$ to the monopole component prior to applying the window convolution \citep{Pinon:2024wzd}\,,
\ba
P_{\ell}(k)=\sum_{\ell^{\prime}} \sum_{k^{\prime}} W_{\ell \ell^{\prime}}\left(k, k^{\prime}\right)\left[P_{\ell^{\prime}}\left(k^{\prime}\right)+N \delta_{\ell^{\prime} 0}\right]\,,
\ea
We determine the value of $N$ by minimizing the residuals between the windowed cubic-box measurements, $\mathbf{W}\cdot\mathbf{P}_{\rm cubic}$, and the average cut-sky multipoles, $\mathbf{P}_{\rm cutsky}$. The fit is restricted to wavenumbers $k < 0.25\,h\,{\rm Mpc}^{-1}$. The solid lines in Figs.~\ref{fig:8} and \ref{fig:9} indicate that allowing the parameter $N$ to vary leads to a significantly improved agreement between the window-convolved cubic prediction and the measured cut-sky power spectra, indicating that residuals can be effectively absorbed by a constant shot-noise-like term.

We next validate the full-shape fitting pipeline using the averaged power spectra from the 25 cut-sky {\tt Abacus-2 complete} mocks. Unlike in the cubic-mock validation, we allow the pre-, post-, and cross-spectra to have free shot-noise-like parameters. These additional degrees of freedom help absorb small residual inaccuracies in the window-matrix convolution and thereby enable a more reliable comparison with the emulator predictions. Each shot-noise-like parameter is assigned a Gaussian prior, $\mathcal{N}\left(0,1^2\right)\times 1/n$.

The results of these fits to the first $z$ bin of {\tt Abacus-2} mocks are shown in the lower block of Fig.\,\ref{fig:mockz1} and in Fig.~\ref{fig:A9} in the Appendix. Both the $\Lambda$CDM and $w$CDM cosmologies are accurately recovered from the cut-sky {\tt Abacus-2 complete} mocks, demonstrating that the window-function implementation and the full-shape pipeline are unbiased at the current statistical precision. Since the redshift ($z=0.8$) of the second $z$ bin of {\tt Abacus-2} differs substantially from that of the emulator ($z=0.689$), we do not perform a mock validation for that bin in this work.

\subsection{DESI DR1 results}

\begin{figure*}[htp]
    \centering
    \includegraphics[width=0.9\textwidth]{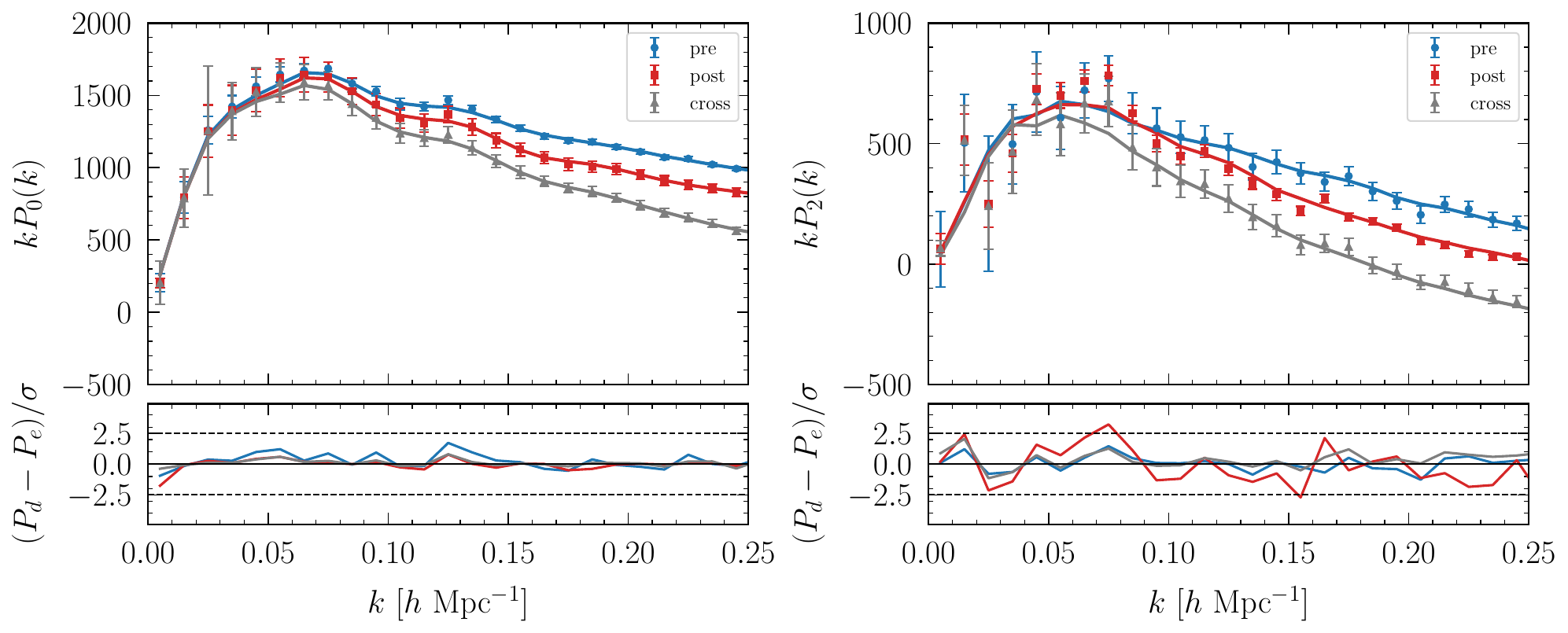}
    \caption{Upper panels: Symbols with error bars show the measured pre-reconstructed, post-reconstructed, and cross power spectrum monopole (left) and quadrupole (right) of the DESI \texttt{LRG1}. The solid curves correspond to the emulator predictions evaluated at the posterior mean values from the CMB+$P_{\rm all}$ constraint on the $\Lambda$CDM model. Lower panels: Residuals between the measured and predicted power spectra, normalized by the total uncertainty. The horizontal dashed lines indicate the $\pm 2.5\sigma$ ranges.}
    \label{fig:11z1}
\end{figure*}

\begin{figure*}[htp]
    \centering
    \includegraphics[width=0.9\textwidth]{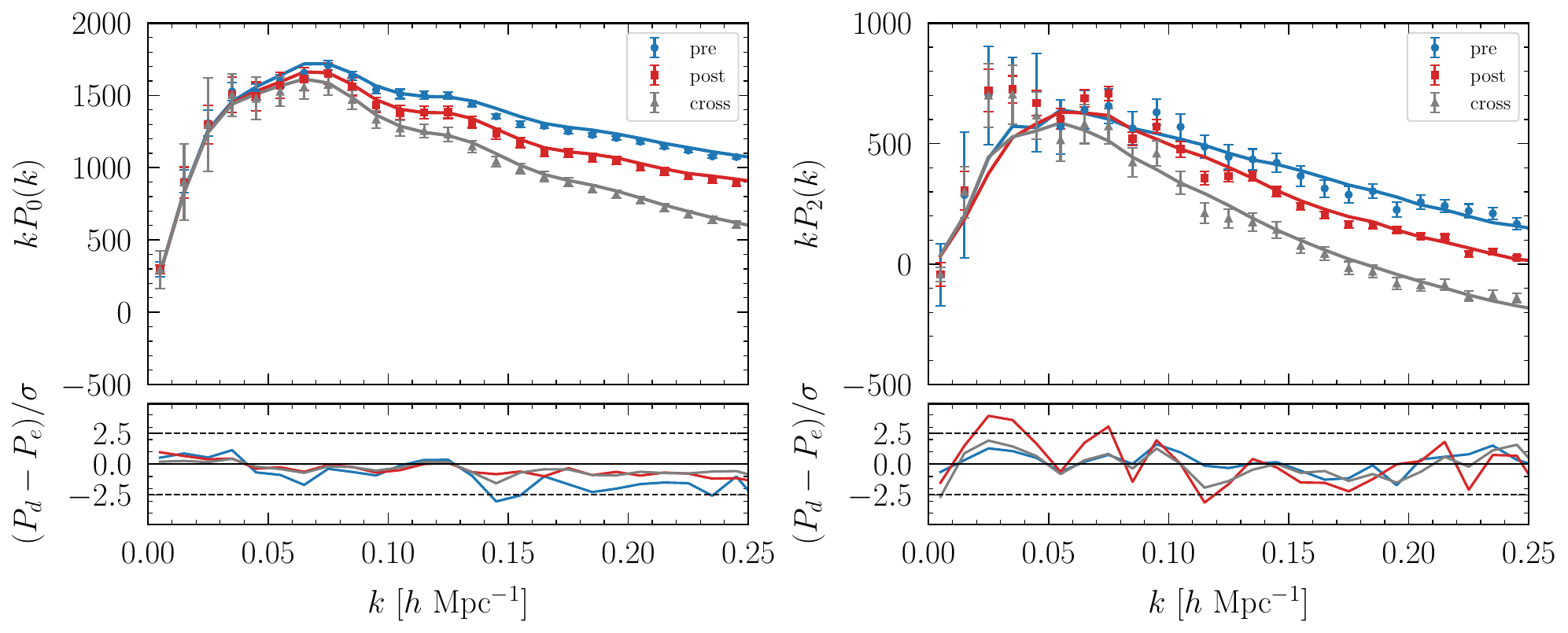}
    \caption{Same as Fig.~\ref{fig:11z1}, but for \texttt{LRG2}.}
    \label{fig:11z2}
\end{figure*}

We employ the DESI DR1 LRG power-spectrum measurements with the ``FA+RIC'' corrections applied, as shown in Figs.~\ref{fig:11z1} and \ref{fig:11z2}. The impact of FA incompleteness and RIC effect on the measured power spectra is shown in Fig.~\ref{fig:A10} in the Appendix. The final power spectrum is obtained by combining the North and South Galactic Cap measurements through a weighted average, where the weights are set by the normalization amplitude of each cap. As indicated by the mock tests, the full-shape information from DR1 alone provides only weak constraints on parameters such as $\Omega_{\mathrm{c}}h^{2}$ and $w$. Therefore, for the cosmological analysis of the real survey data, we combine the DESI full-shape likelihood with external datasets, specifically the CMB distance priors and SN Ia datasets.

\begin{figure*}[htp]
    \centering
    \includegraphics[width=0.45\textwidth]{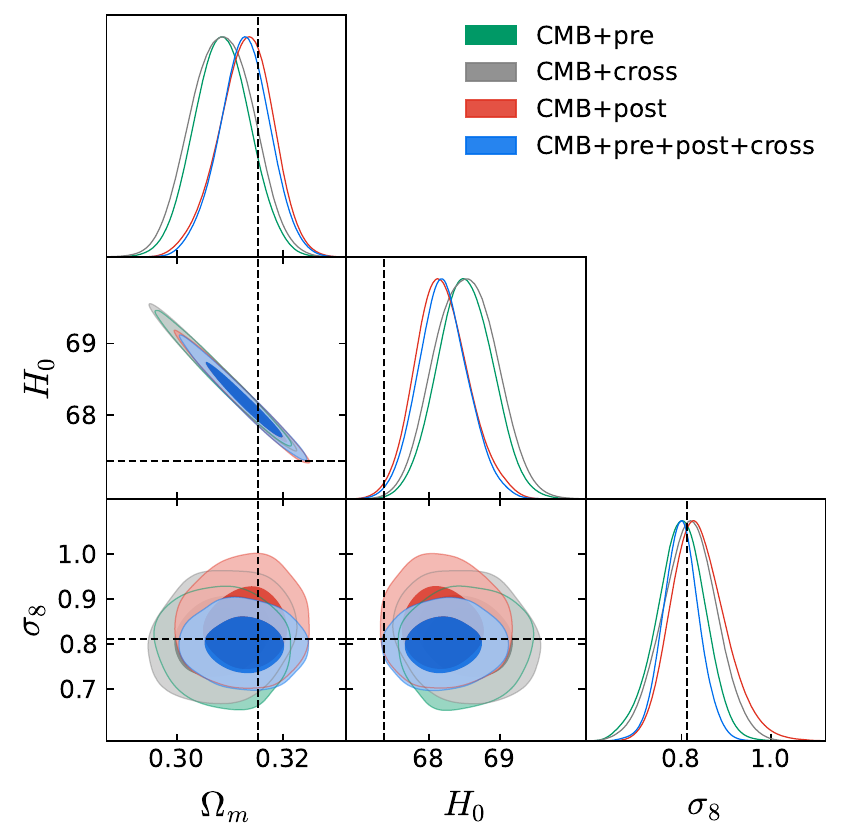}
    \includegraphics[width=0.45\textwidth]{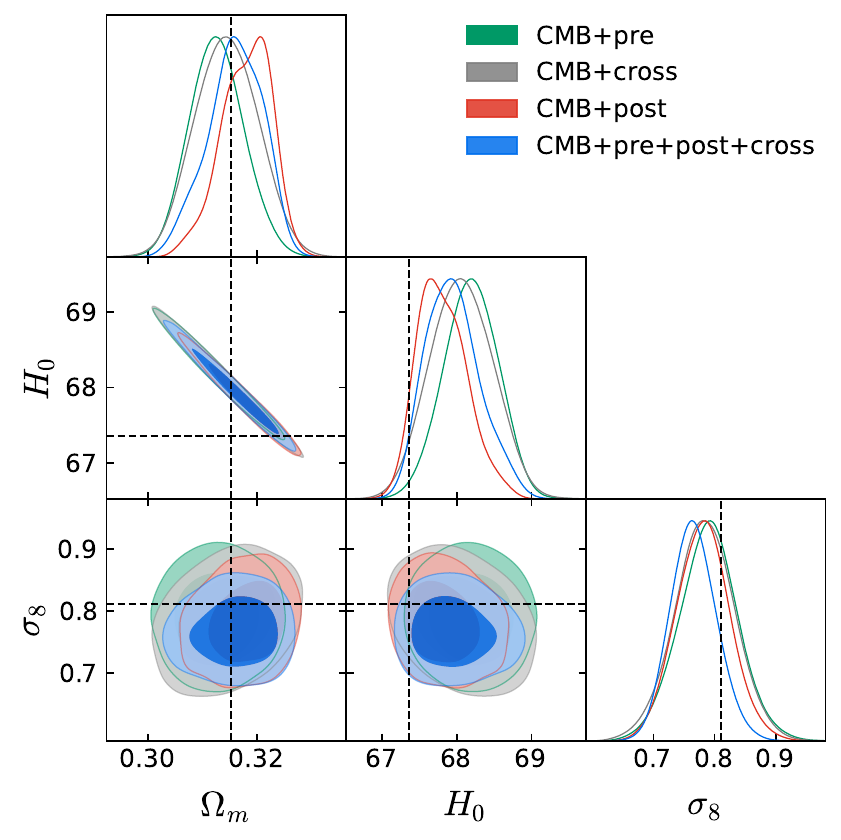}
    \caption{Constraints on the $\Lambda$CDM cosmological model from the combination of CMB distance priors and $P_{\rm T}$ ($T={\mathrm{pre},\mathrm{post},\mathrm{cross},\mathrm{all}}$). Left: \texttt{LRG1}; Right: \texttt{LRG2}. The progressively tighter contours when going from $P_{\rm pre}$ to $P_{\rm all}$ illustrate the additional information gained by including $P_{\rm post}$ and $P_{\rm cross}$. The dashed lines indicate the \textit{Planck} 2018 $\Lambda$CDM cosmology.}
    \label{fig:12}
\end{figure*}

We first consider constraints on the $\Lambda$CDM parameters. For each of the power-spectrum combinations $P_{\rm T}$ ($\rm T={\mathrm{pre},\mathrm{post},\mathrm{cross},\mathrm{all}}$), we combine the full-shape likelihood with the CMB distance information. The resulting posteriors are shown in the triangle plot of Fig.~\ref{fig:12}. Relative to using the pre-reconstruction spectrum alone \footnote{We adopt the emulator-based pre-reconstructed results as our baseline and quantify the improvement relative to that reference. A detailed comparison with the perturbation-theory-based (PT-based) analysis in \citep{DESI:2024jxi} is left to future work, as the two approaches differ in modeling frameworks, parameterizations, and prior choices, making a direct comparison non-trivial.}, the joint fit using all three spectra, $P_{\rm all}$, yields an improvement of $20$-$25\%$ in the constraint on $\sigma_8$, demonstrating the additional nonlinear information encoded in the combined pre-post-cross $P_{\ell}(k)$ analysis. As summarized in Table~\ref{tab:result}, the multi-parameter Figure of Merit (FoM) \citep{Wang:2008zh} in the $(\Omega_m, H_0, \sigma_8)$ subspace also increases when including $P_{\rm post}$ and $P_{\rm cross}$, with FoM ratios of $1.27$ and $1.46$ for \texttt{LRG1} and \texttt{LRG2}, respectively, compared to the CMB+$P_{\rm pre}$ baseline.

\begin{figure*}[htp]
    \centering
    \includegraphics[width=0.45\textwidth]{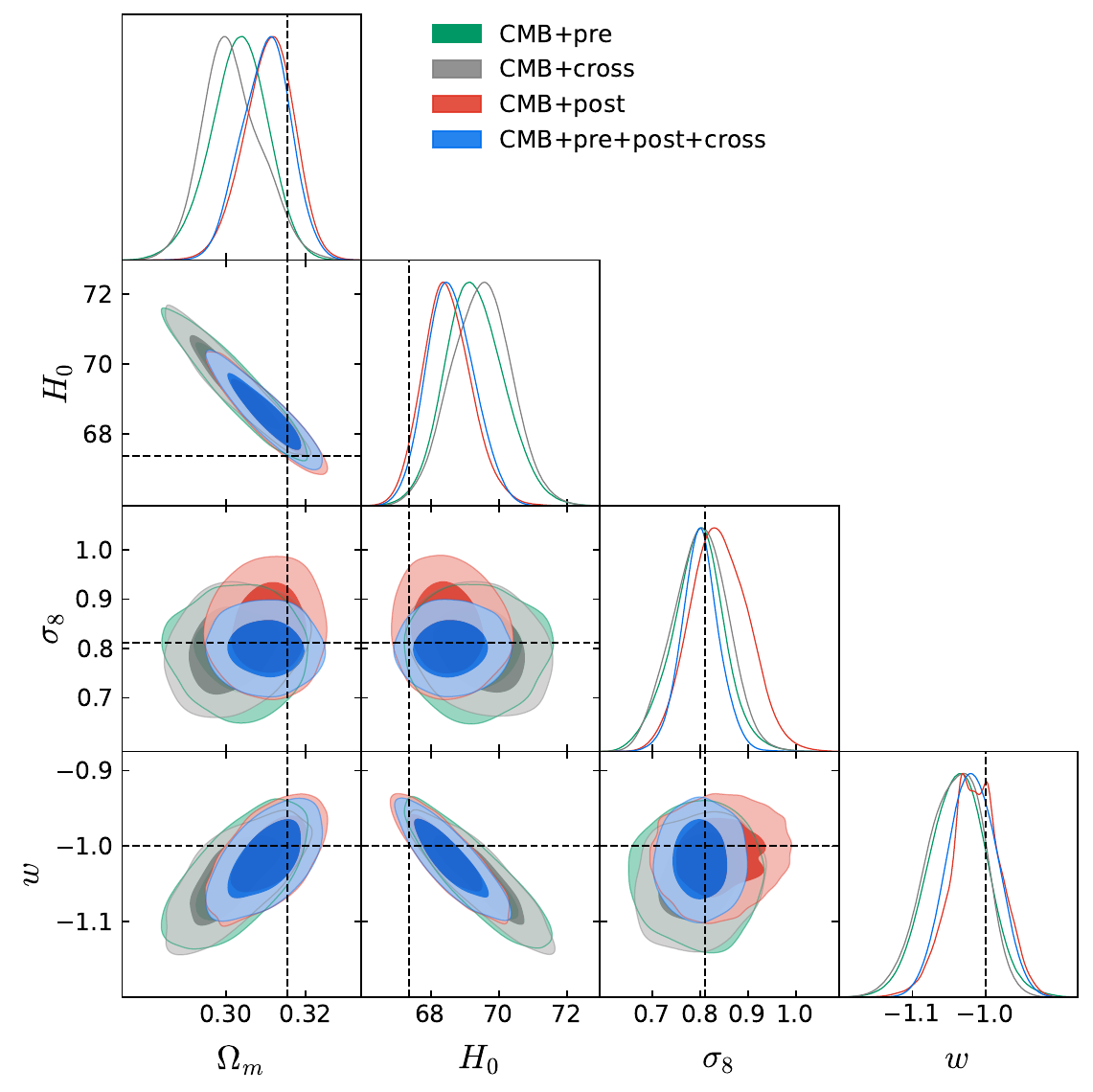}
    \includegraphics[width=0.45\textwidth]{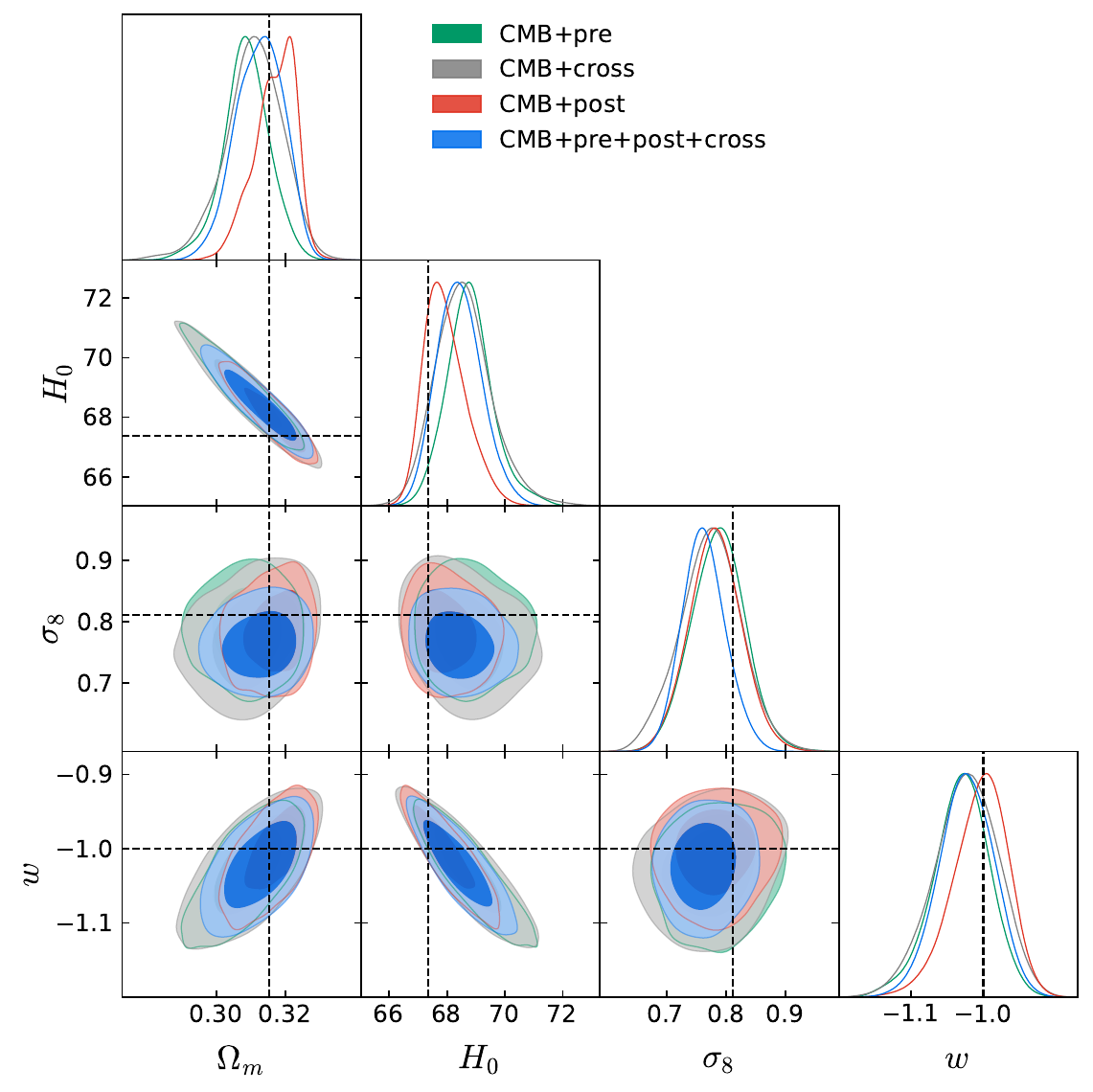}
    \caption{Same as Fig.~\ref{fig:12}, but for the $w$CDM cosmological model. The inclusion of $P_{\rm post}$ and $P_{\rm cross}$ again leads to visibly tighter constraints, particularly in the $(\Omega_m,\sigma_8)$ and $(w,\sigma_8)$ planes.}
    \label{fig:13}
\end{figure*}

We then extend the analysis to the $w$CDM model. The constraints obtained from CMB+$P_{\rm T}$ are presented in Fig.~\ref{fig:13}. Again, the combination $P_{\rm all}$ delivers the tightest constraints, with the marginalized error on $\sigma_8$ reduced by $18$-$27\%$ relative to CMB+$P_{\rm pre}$ across the two redshift bins and the FoM in the $(\Omega_m, H_0,\sigma_8)$ subspace enhanced by factors of $1.81$ and $1.18$ for \texttt{LRG1} and \texttt{LRG2}, respectively (Table~\ref{tab:result}). The mean values of $w$ remain close to $-1$ in all cases, indicating consistency with a cosmological constant.

\begin{figure*}[htp]
    \centering
    \includegraphics[width=0.9\textwidth]{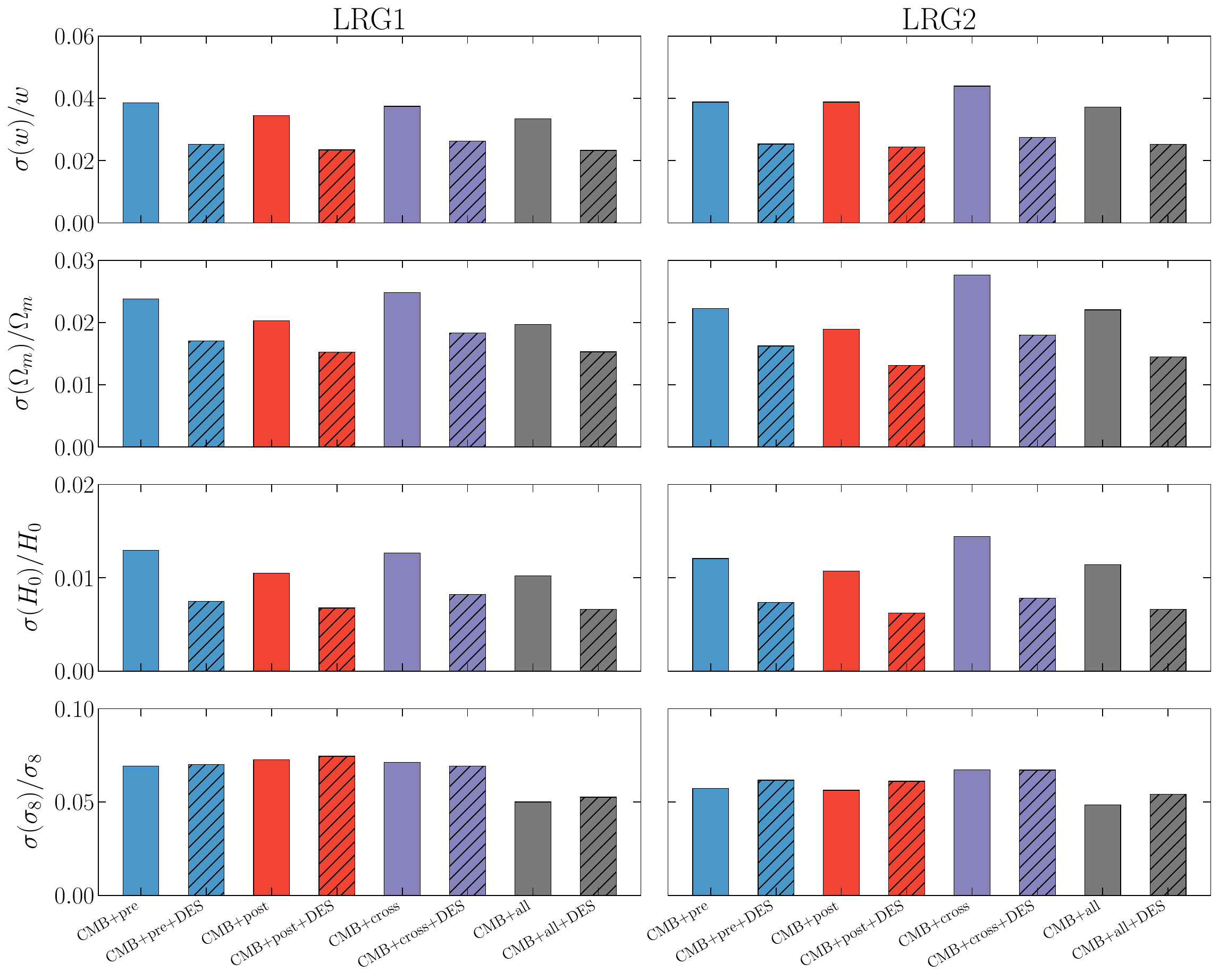}
    \caption{Fractional uncertainties on the cosmological parameters $(w,\Omega_m,H_0,\sigma_8)$ in the $w$CDM model for the two DESI LRG redshift bins (left: \texttt{LRG1}, right: \texttt{LRG2}) under various data combinations. The x-axis shows combinations of the CMB distance prior with individual power-spectrum components ($P_{\rm pre}$, $P_{\rm post}$, $P_{\rm cross}$) and their full combination ($P_{\rm all}$). For each case, results without DES-Dovekie supernovae are shown in solid colors, while the corresponding combinations including the DES-Dovekie SN sample are indicated by hatched bars.}
    \label{fig:wCDM}
\end{figure*}

Finally, we present the $w$CDM constraints obtained by combining various power-spectrum components with the CMB distance prior and each of the three Type~Ia supernova samples, \ie~Union3, PantheonPlus, and DES-Dovekie, as summarized in Table~\ref{tab:wCDM} in the Appendix. The results from the three SN Ia datasets are highly consistent with one another. When comparing constraints derived from each individual power-spectrum component with those from the full combination, $P_{\rm all}$ consistently provides the tightest limits. Among the three supernova samples, the DES-Dovekie combination yields the strongest overall constraints. Fractional uncertainties on $(w, \Omega_m, H_0, \sigma_8)$ are shown in Fig.~\ref{fig:wCDM}, illustrating how the inclusion of post-reconstruction and cross power spectra, together with DES-Dovekie supernova data, progressively improves the parameter constraints. Relative to the case without supernovae (\ie, CMB+$P_{\rm T}$), adding DES-Dovekie supernova data further tightens the constraints on the background parameters $(w, \Omega_m, H_0)$, while having minimal impact on $\sigma_8$.

For the \texttt{LRG1} sample, the joint fits in $w$CDM yield
\ba
\left.\begin{array}{rl}
\Omega_{\mathrm{m}} & =0.314 \pm 0.0051, \\
w & =-0.987 \pm 0.027,
\end{array}\right\} \begin{aligned}
& \text { $P_{\rm all}$ }+ \text { CMB } \\
& \text { + Union3, }
\end{aligned}
\ea
\ba
\left.\begin{array}{rl}
\Omega_{\mathrm{m}} & =0.313 \pm 0.0049, \\
w & =-0.992 \pm 0.024,
\end{array}\right\} \begin{aligned}
& \text { $P_{\rm all}$ }+ \text { CMB } \\
& \text { + PantheonPlus, }
\end{aligned}
\ea
and
\ba
\left.\begin{array}{rl}
\Omega_{\mathrm{m}} & =0.314 \pm 0.0048, \\
w & =-0.988 \pm 0.023,
\end{array}\right\} \begin{aligned}
& \text { $P_{\rm all}$ }+ \text { CMB } \\
& \text { + DES-Dovekie.}
\end{aligned}
\ea

\begin{figure*}[htp]
    \centering
    \includegraphics[width=0.45\textwidth]{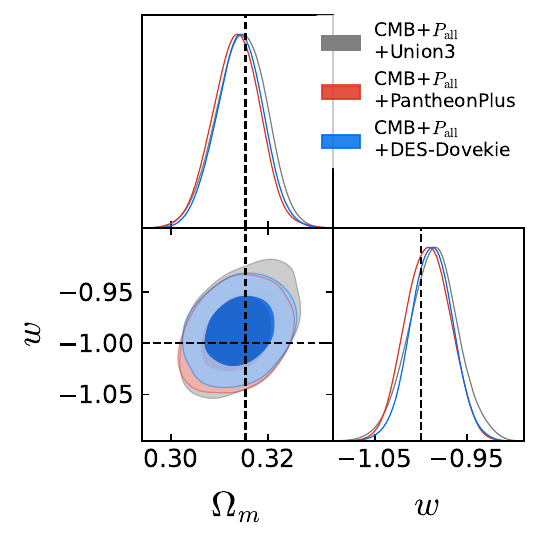}
    \includegraphics[width=0.45\textwidth]{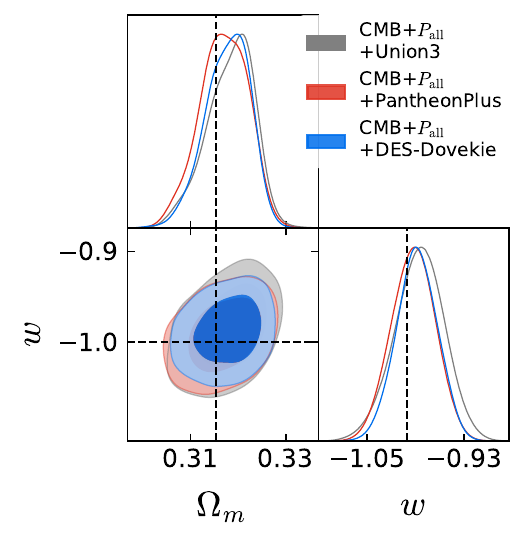}
    \caption{Constraints on the $w$CDM cosmological model from the combination of CMB distance priors, $P_{\rm all}$, and different Type~Ia supernova datasets (PantheonPlus, Union3, and DES-Dovekie). Left: \texttt{LRG1}; Right: \texttt{LRG2}. The dashed lines indicate the \textit{Planck} 2018 $\Lambda$CDM cosmology. The three SN compilations yield mutually consistent constraints on $(\Omega_m, w)$, all compatible with $w\simeq -1$.}
    \label{fig:14}
\end{figure*}

Similarly, for the \texttt{LRG2} sample, we find
\ba
\left.\begin{array}{rl}
\Omega_{\mathrm{m}} & =0.318 \pm 0.0050, \\
w & =-0.985 \pm 0.031,
\end{array}\right\} \begin{aligned}
& \text { $P_{\rm all}$ }+ \text { CMB } \\
& \text { + Union3,}
\end{aligned}
\ea
\ba
\left.\begin{array}{rl}
\Omega_{\mathrm{m}} & = 0.317 \pm 0.0051, \\
w & =-0.993 \pm 0.027,
\end{array}\right\} \begin{aligned}
& \text { $P_{\rm all}$ }+ \text { CMB } \\
& \text { + PantheonPlus, }
\end{aligned}
\ea
and
\ba
\left.\begin{array}{rl}
\Omega_{\mathrm{m}} & = 0.318 \pm 0.0046, \\
w & =-0.988 \pm 0.025,
\end{array}\right\} \begin{aligned}
& \text { $P_{\rm all}$ }+ \text { CMB } \\
& \text { + DES-Dovekie.}
\end{aligned}
\ea
The results obtained from the three SN~Ia compilations are mutually consistent within $1\sigma$, as shown in Fig.~\ref{fig:14}, and all combinations favour a value of $w$ close to $-1$. Compared to the constraints on $\Omega_m$ and $w$ obtained without SNe~Ia, this demonstrates that our nonlinear-information-enhanced DESI constraints are in good agreement with independent probes of the late-time expansion history provided by SNe~Ia.

\begin{table*}[!t]
\centering
\renewcommand{\arraystretch}{1.3}
\resizebox{\textwidth}{!}{%
\begin{tabular}{c|cccc|cccc}
\hline\hline
           & \multicolumn{4}{c|}{\texttt{LRG1} } & \multicolumn{4}{c}{\texttt{LRG2} } \\
dataset: CMB+    &  $P_{\rm pre}$ & $ P_{\rm post}$ & $P_{\rm cross}$  & $P_{\rm all}$ &  $P_{\rm pre}$ & $ P_{\rm post}$ & $P_{\rm cross}$  & $P_{\rm all}$\\\hline
\rule{0pt}{4ex}{$\Lambda$CDM} & & &&&&&&\\
$\omega_c$ & $0.12170 \pm 0.00092$ & $0.12241 \pm 0.00088$ & $0.12169 \pm 0.00100$& $0.12238 \pm 0.00085$ & $0.12236 \pm 0.00087$& $0.12326 \pm 0.00080$&$0.12266 \pm 0.00098$& $0.12289 \pm 0.00089$\\
$\Omega_{de}$ & $0.6914 \pm 0.0052$  & $0.6871 \pm 0.0051$ &$0.6915 \pm 0.0058$&$0.6873 \pm 0.0048$& $0.6873 \pm 0.0050$&$0.6820 \pm 0.0046$&$0.6855 \pm 0.0057$&$0.6840 \pm 0.0051$\\
$\ln(10^{10}A_s)$   & $2.980 \pm 0.141$ & $3.076 \pm 0.144$ & $3.030 \pm 0.150$&$2.980 \pm 0.104$&$2.957 \pm 0.123$&$2.929 \pm 0.112$&$2.939 \pm 0.125$& $2.893 \pm 0.099$\\
\rule{0pt}{4ex}$\rm FoM$ [$\omega_c, \Omega_{de},\ln(10^{10}A_s)$]& $1$ &$1$ &$0.89$ &$1.47$ &$1$ &$1.17$ &$0.89$ &$1.23$ \\\hline
$\Omega_m$ & $0.3086 \pm 0.0052$& $0.3129 \pm 0.0051$ & $0.3095 \pm 0.0058$ &$0.3127 \pm 0.0048$& $0.3127 \pm 0.0050$ &$0.3180 \pm 0.0046$&$0.3145 \pm 0.0057$&$0.3160 \pm 0.0051$\\ 
$H_0$  & $68.51 \pm 0.39 $ &$68.19 \pm 0.38 $&$68.51 \pm 0.43 $ &$68.20 \pm 0.35$& $68.19 \pm 0.36$ &$67.81 \pm 0.34$ &$68.06 \pm 0.42$  &$67.95 \pm 0.37$\\ 
$\sigma_8$  &$0.798 \pm 0.055 $ &$0.839 \pm 0.061 $&$0.818 \pm 0.061 $&$0.799 \pm 0.041 $& $0.790 \pm 0.048$&$0.781 \pm 0.044 $&$0.784 \pm 0.049 $&$0.766 \pm 0.038 $\\ 
\rule{0pt}{4ex}$\rm FoM$ [$\Omega_m, H_0,\sigma_8$]& $1$ &$0.95$ &$0.86$ &$1.46$ &$1$ &$1.19$ &$0.89$ &$1.27$ \\
\hline\hline
\rule{0pt}{4ex}{$w$CDM} & & &&&&&&\\
$\omega_c$ & $0.12250 \pm 0.00138$ & $0.12267 \pm 0.00132$& $0.12242 \pm 0.00130$&$0.12276 \pm 0.00129$ & $0.12305 \pm 0.00134$ & $0.12326 \pm 0.00139$&$0.12309 \pm 0.00151$& $0.12335 \pm 0.00144$\\
$\Omega_{de}$ & $0.6968 \pm 0.0072$ & $0.6893 \pm 0.0063$ &$0.6984 \pm 0.0075$&$0.6900 \pm 0.0061$&$0.6913 \pm 0.0069$&$0.6831 \pm 0.0060$&$0.6892 \pm 0.0086$& $0.6874 \pm 0.0069$\\
$\ln(10^{10}A_s)$   & $2.943 \pm 0.145$ & $3.068 \pm 0.150 $&$2.952 \pm 0.152$& $2.970 \pm 0.105$&$2.925 \pm 0.125$&$2.930 \pm 0.119$&$2.895 \pm 0.145$&$2.867 \pm 0.104$\\
$w$   &$-1.038 \pm 0.040 $&$-1.014 \pm 0.035 $&$-1.042 \pm 0.039 $&$-1.018 \pm 0.034 $&$-1.031 \pm 0.040 $&$-1.005 \pm 0.039$&$-1.024 \pm 0.045$&$-1.023 \pm 0.038$\\ 
\rule{0pt}{4ex}$\rm FoM$ [$\omega_c, \Omega_{de},\ln(10^{10}A_s),w$]& $1$ &$1.26$ &$1.08$ &$1.87$ &$1$ &$1.17$ &$0.64$ &$1.19$ \\\hline
$\Omega_m$  & $0.3032 \pm 0.0072$&$0.3107 \pm 0.0063$&$0.3016 \pm 0.0075 $&$0.3100 \pm 0.0061 $&$0.3087 \pm 0.0069 $&$0.3169 \pm 0.0060 $  &$0.3108 \pm 0.0086 $ &$0.3126 \pm 0.0069 $ \\ 
$H_0$  & $67.30 \pm 0.87 $ &$68.49 \pm 0.72$ &$69.47 \pm 0.88 $ &$68.59 \pm 0.70 $&$68.79 \pm 0.83 $&$67.94 \pm 0.73 $&$68.57 \pm 0.99 $&$68.43 \pm 0.78 $\\ 
$\sigma_8$  &$0.795 \pm 0.055 $&$ 0.840 \pm0.061$&$0.800 \pm 0.057 $&$0.800 \pm 0.040 $&$0.787 \pm 0.045 $&$0.782 \pm 0.044 $ &$0.774 \pm 0.052 $ &$0.763 \pm 0.037 $\\ 
\rule{0pt}{4ex}$\rm FoM$ [$\Omega_m, H_0,\sigma_8$]  &$1$ &$1.13$ &$1.03$ &$1.81$ &$1$ &$1.16$ &$0.65$ &$1.18$\\ 
\hline\hline
\end{tabular}
}
\renewcommand{\arraystretch}{1}
\caption{Summary of constraints on the basic cosmological parameters [$\omega_c, \Omega_{de},\ln(10^{10}A_s),w$] and derived cosmological parameters  ($\Omega_m, H_0,\sigma_8$) in $\Lambda$CDM model and $w$CDM model from the combination of CMB distance priors with different choices of power-spectrum statistics $P_T$ for the \texttt{LRG1} and \texttt{LRG2} samples. The Figure of Merit (FoM) values are normalized to the CMB+$P_{\rm pre}$ case in each model.}
\label{tab:result}
\end{table*}

\section{Conclusion} \label{sec:summary}

In this work, we have developed and applied an emulator-based framework to jointly analyse the pre- and post-reconstruction power spectra, $P_{\rm pre}$ and $P_{\rm post}$, together with their cross-power spectrum, $P_{\rm cross}$, for the DESI DR1 Luminous Red Galaxy sample. Using the \textsc{Dark Quest} simulation suite to build an emulator that accurately models the reconstruction process and its impact on the full shape, we validated our pipeline with a comprehensive set of cubic and cut-sky mocks, including \textsc{Dark Quest}, {\tt Abacus-2}, and {\tt Abacus-HF}, and showed that it reliably recovers the input cosmology under a variety of controlled conditions.

Applied to the DESI DR1 \texttt{LRG1} and \texttt{LRG2} samples, our method demonstrates that combining $P_{\rm pre}$, $P_{\rm post}$, and $P_{\rm cross}$ allows one to harvest additional nonlinear information from the reconstructed density fields. When the DESI full-shape likelihood is combined with CMB distance priors, the joint analysis $P_{\rm all}$ improves the $\sigma_8$ constraint by $\simeq 20$-$25\%$ relative to the CMB+$P_{\rm pre}$ baseline in $\Lambda$CDM, and enhances the multi-parameter Figure of Merit in the $(\Omega_m,H_0,\sigma_8)$ subspace by $27$-$46\%$, depending on the redshift bin. Extending the model to a $w$CDM framework with a constant dark-energy equation-of-state parameter, we find that the joint CMB+$P_{\rm all}$ combination improves the constraints on $\sigma_8$, $w$ and $H_0$ by approximately $17$-$27\%$, $5$-$15\%$, and $6$-$20\%$, respectively, across the two LRG redshift bins, relative to using only CMB+$P_{\rm pre}$. The inclusion of three SN~Ia datasets (\ie, Union3, PantheonPlus, and DES-Dovekie) respectively further tightens the constraints on the background cosmological parameters, with all combinations yielding values of $w$ consistent with a cosmological constant and mutually compatible at the $\lesssim 1\sigma$ level.

These results provide a concrete proof of concept that pre- and post-reconstruction power spectra, together with their cross-correlation, can be robustly modelled and used to extract nonlinear information from current-generation spectroscopic surveys. Looking ahead to DESI DR2 and beyond, the rapidly shrinking statistical errors will demand a corresponding improvement in the precision and accuracy of the emulator. This will require denser sampling of the cosmological and HOD parameter space, refined treatments of observational effects, and tighter control of emulator-systematics so that theoretical uncertainties remain subdominant to the data. With such improvements, the methodology presented here can be straightforwardly applied to the larger DR2 LRG samples and extended in parameter space, e.g., to models with evolving dark energy or massive neutrinos.

Finally, the framework we have developed is not limited to LRGs. It can be applied to other DESI tracers such as emission line galaxies, bright galaxy samples, and quasars, as well as to multi-tracer combinations that exploit cross-correlations between different populations. Extending the emulator to these tracers will enable a coherent, survey-wide extraction of nonlinear information from DESI, and similar analyses can be envisaged for forthcoming surveys such as Euclid and 4MOST. Our results thus represent an important step toward fully realizing the potential of reconstruction-enhanced full-shape measurements as precision probes of structure growth and dark energy.

\section{Data Availability}
The data including measured power spectra, window functions, and covariance matrices used in this analysis will be available in a Zenodo repository after publication.

{\it Acknowledgments}—This work is supported by National Key R\&D Program of China No. (2023YFA1607800, 2023YFA1607803), NSFC Grants (12273048, 12422301), the CAS Project for Young Scientists in Basic Research (No. YSBR-092), the Youth Innovation Promotion Association CAS., and the New Cornerstone Science Foundation through the XPLORER prize. KK is supported by STFC grant number ST/W001225/1 and ST/B001175/1. YW acknowledges the use of the UK Sciama High Performance Computing cluster, supported by the Institute of Cosmology and Gravitation (ICG), University of Portsmouth.

This material is based upon work supported by the U.S. Department of Energy (DOE), Office of Science, Office of High-Energy Physics, under Contract No. DE–AC02–05CH11231, and by the National Energy Research Scientific Computing Center, a DOE Office of Science User Facility under the same contract. Additional support for DESI was provided by the U.S. National Science Foundation (NSF), Division of Astronomical Sciences under Contract No. AST-0950945 to the NSF’s National Optical-Infrared Astronomy Research Laboratory; the Science and Technology Facilities Council of the United Kingdom; the Gordon and Betty Moore Foundation; the Heising-Simons Foundation; the French Alternative Energies and Atomic Energy Commission (CEA); the National Council of Humanities, Science and Technology of Mexico (CONAHCYT); the Ministry of Science, Innovation and Universities of Spain (MICIU/AEI/10.13039/501100011033), and by the DESI Member Institutions: \url{https://www.desi.lbl.gov/collaborating-institutions}.

The DESI Legacy Imaging Surveys consist of three individual and complementary projects: the Dark Energy Camera Legacy Survey (DECaLS), the Beijing-Arizona Sky Survey (BASS), and the Mayall z-band Legacy Survey (MzLS). DECaLS, BASS and MzLS together include data obtained, respectively, at the Blanco telescope, Cerro Tololo Inter-American Observatory, NSF’s NOIRLab; the Bok telescope, Steward Observatory, University of Arizona; and the Mayall telescope, Kitt Peak National Observatory, NOIRLab. NOIRLab is operated by the Association of Universities for Research in Astronomy (AURA) under a cooperative agreement with the National Science Foundation. Pipeline processing and analyses of the data were supported by NOIRLab and the Lawrence Berkeley National Laboratory. Legacy Surveys also uses data products from the Near-Earth Object Wide-field Infrared Survey Explorer (NEOWISE), a project of the Jet Propulsion Laboratory/California Institute of Technology, funded by the National Aeronautics and Space Administration. Legacy Surveys was supported by: the Director, Office of Science, Office of High Energy Physics of the U.S. Department of Energy; the National Energy Research Scientific Computing Center, a DOE Office of Science User Facility; the U.S. National Science Foundation, Division of Astronomical Sciences; the National Astronomical Observatories of China, the Chinese Academy of Sciences and the Chinese National Natural Science Foundation. LBNL is managed by the Regents of the University of California under contract to the U.S. Department of Energy. The complete acknowledgments can be found at \url{https://www.legacysurvey.org/}.

Any opinions, findings, and conclusions or recommendations expressed in this material are those of the author(s) and do not necessarily reflect the views of the U. S. National Science Foundation, the U. S. Department of Energy, or any of the listed funding agencies.

The authors are honored to be permitted to conduct scientific research on I'oligam Du'ag (Kitt Peak), a mountain with particular significance to the Tohono O’odham Nation.

\bibliographystyle{JHEP}
\bibliography{draft}

\newpage

\begin{center}
\begin{Large}
{\bf Appendix}
\end{Large}
\end{center}

\begin{figure*}[htp]
    \centering
    \includegraphics[width=0.9\textwidth]{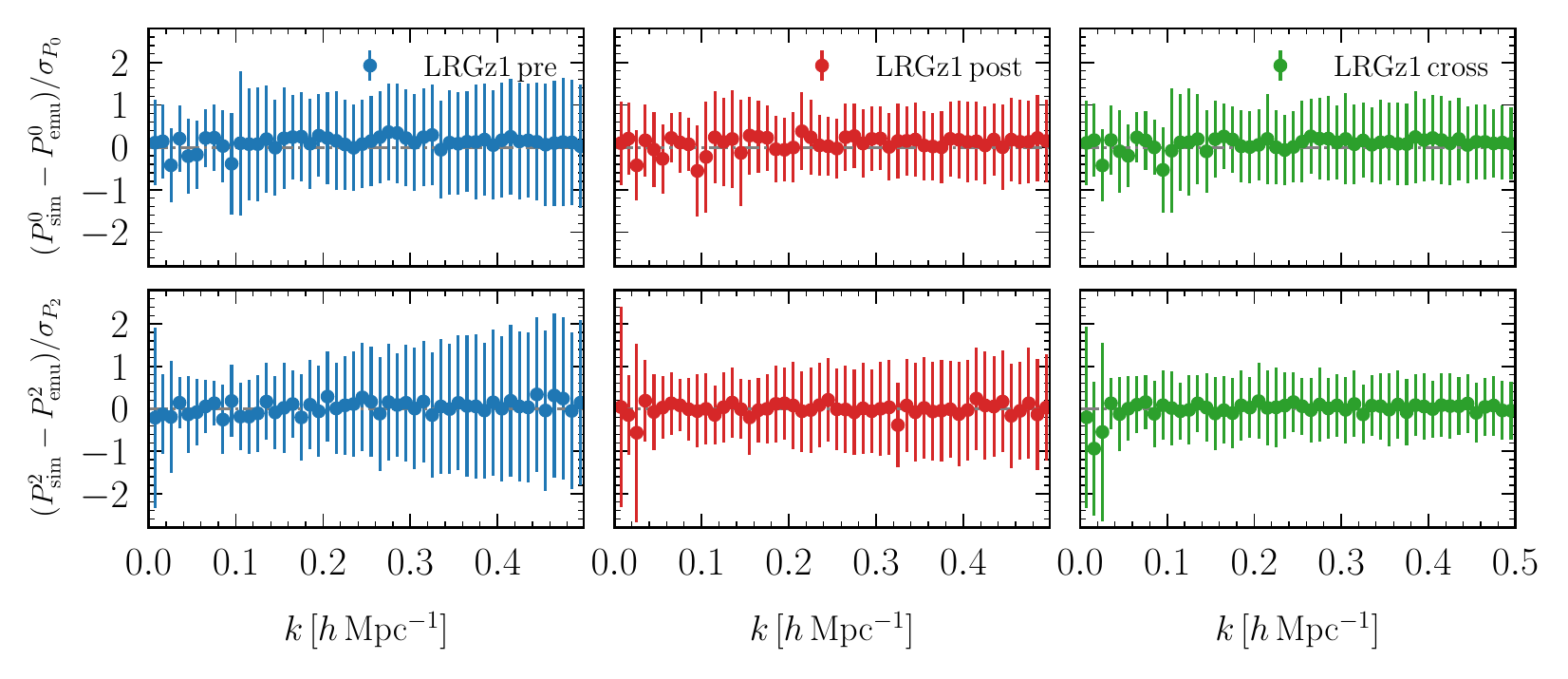}
    \caption{Residuals between the emulated and measured power spectrum monopole (upper panels) and quadrupole (lower panels) from the testing set, normalized by the statistical uncertainty. For the $P_{\rm pre}$ quadrupole, the emulator accuracy gradually exceeds the $1\,\sigma$ statistical uncertainty at scales beyond $k \sim 0.3\,h\,\mathrm{Mpc}^{-1}$. In contrast, both $P_{\rm post}$ and $P_{\rm cross}$ exhibit reduced residual scatter and improved error precision over the scale range up to $k =0.5~h\,\mathrm{Mpc}^{-1}$.}
    \label{fig:A1}
\end{figure*}

\begin{figure*}[htp]
    \centering
    \includegraphics[width=0.9\textwidth]{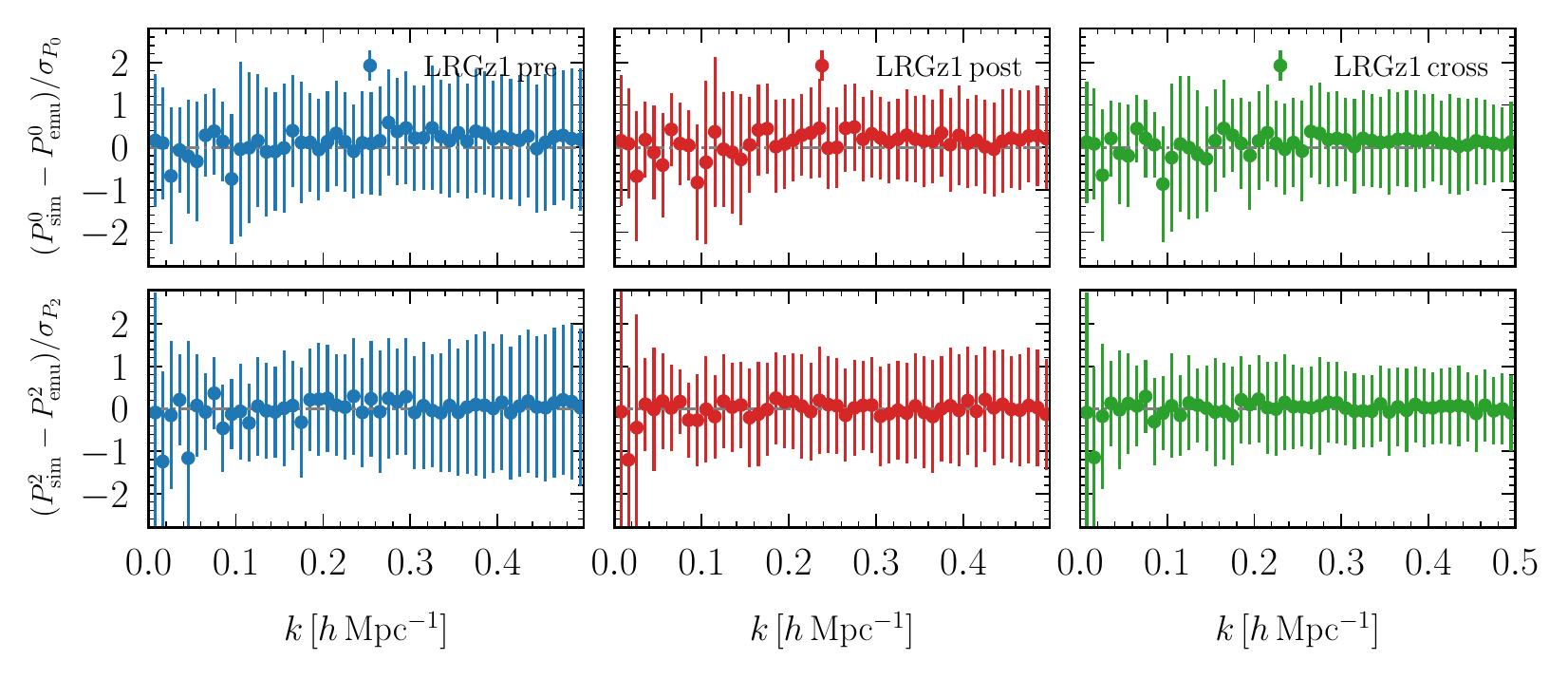}
    \caption{Same as \ref{fig:A1}, but for LRG$z2$.}
    \label{fig:A2}
\end{figure*}

\begin{figure*}[htp]
    \centering
    \includegraphics[width=0.9\textwidth]{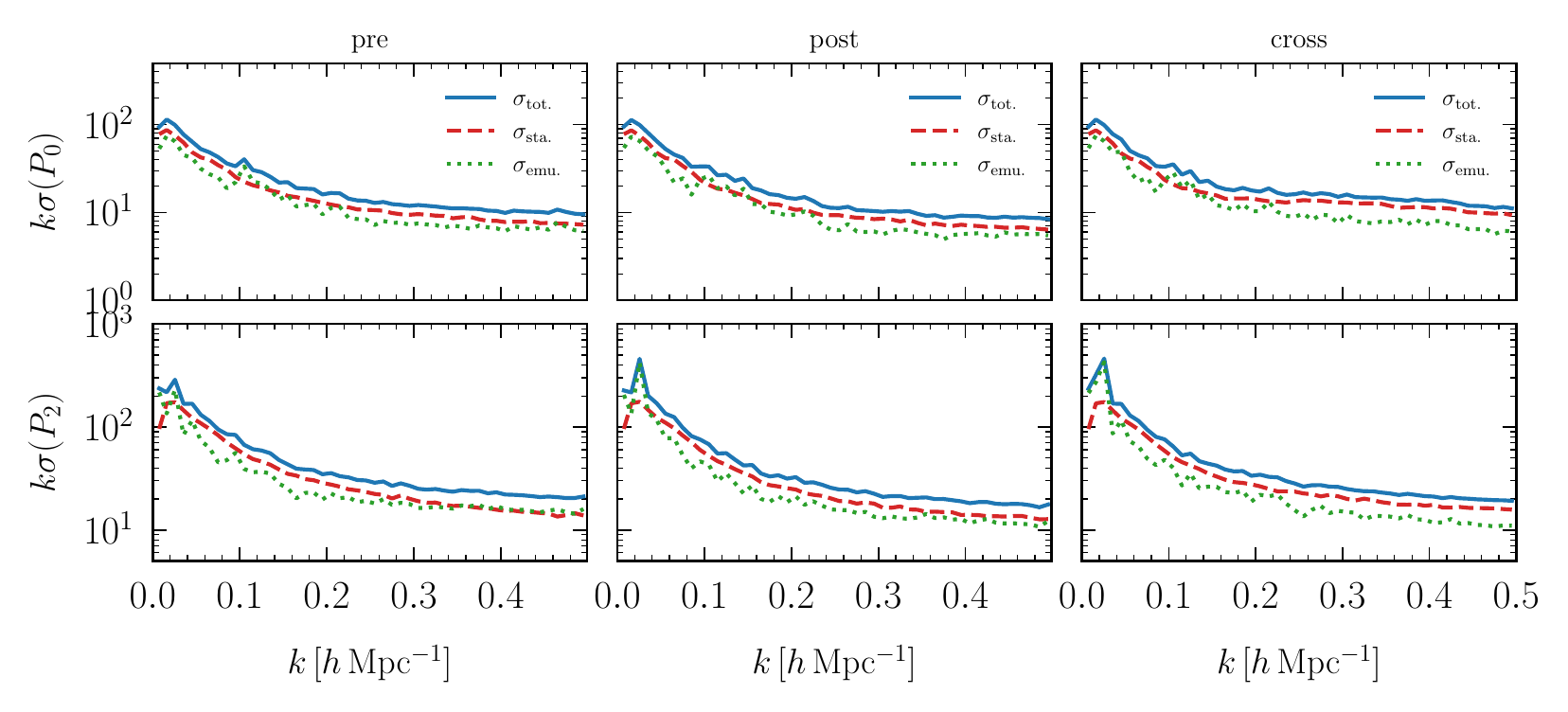}
    \caption{Estimated uncertainties of the power spectrum monopole (upper panels) and quadrupole (lower panels) for the pre-reconstructed field, post-reconstructed field, and their cross between two fields (from left to right). The total uncertainty, $\sigma_{\rm tot.}$ (blue solid), is decomposed into the statistical component, $\sigma_{\rm sta.}$ (red dashed), and the emulator uncertainty, $\sigma_{\rm emu.}$ (green dotted).}
    \label{fig:A3}
\end{figure*}

\begin{figure*}[htp]
    \centering
    \includegraphics[width=0.9\textwidth]{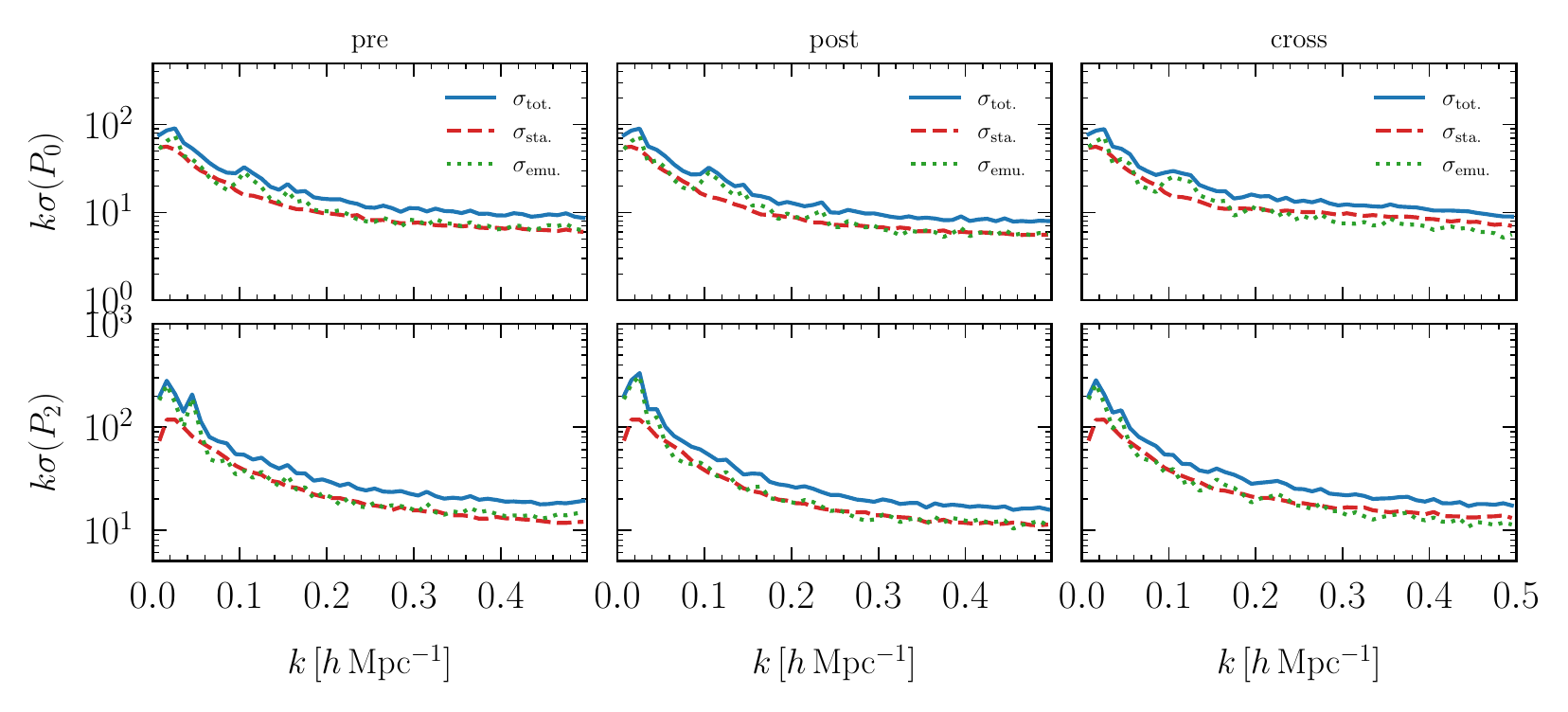}
    \caption{Same as \ref{fig:A3}, but for LRG$z2$.}
    \label{fig:A4}
\end{figure*}

We assess the accuracy of our emulator using $1000$ testing mocks that were not part of the training set. The results are shown in Figs.~\ref{fig:A1}-\ref{fig:A4}. Figs.~\ref{fig:A1} and Figs.~\ref{fig:A2} show the residuals between the emulated and measured quadrupole power spectra, normalized by the statistical uncertainty, in both redshift bins. These results demonstrate that the emulator reproduces the simulated power spectra to within statistical precision over a wide range of wavenumbers, depending the type of power spectra. Compared to $P_{\rm pre}$, both $P_{\rm post}$ and $P_{\rm cross}$ exhibit reduced residual scatter and improved error precision over a wide range of scales. Figs.~\ref{fig:A3} and Figs.~\ref{fig:A4} present the estimated uncertainties for the power spectrum monopole and quadrupole for the pre-reconstructed field, post-reconstructed field, and their cross power spectra in two redshift bins. The total uncertainty, $\sigma_{\rm tot.} = \sqrt{\sigma^2_{\rm sta.} + \sigma^2_{\rm emu.}}$, is the sum of the statistical error, $\sigma_{\rm sta.}$, and the emulator error, $\sigma_{\rm emu.}$. The emulator for $P_{\rm post}$ and $P_{\rm cross}$ maintains accuracy at the level of the statistical uncertainty over the scale range up to $k =0.5~h\,\mathrm{Mpc}^{-1}$.

\begin{figure*}[htp]
    \centering
    \includegraphics[width=0.45\textwidth]{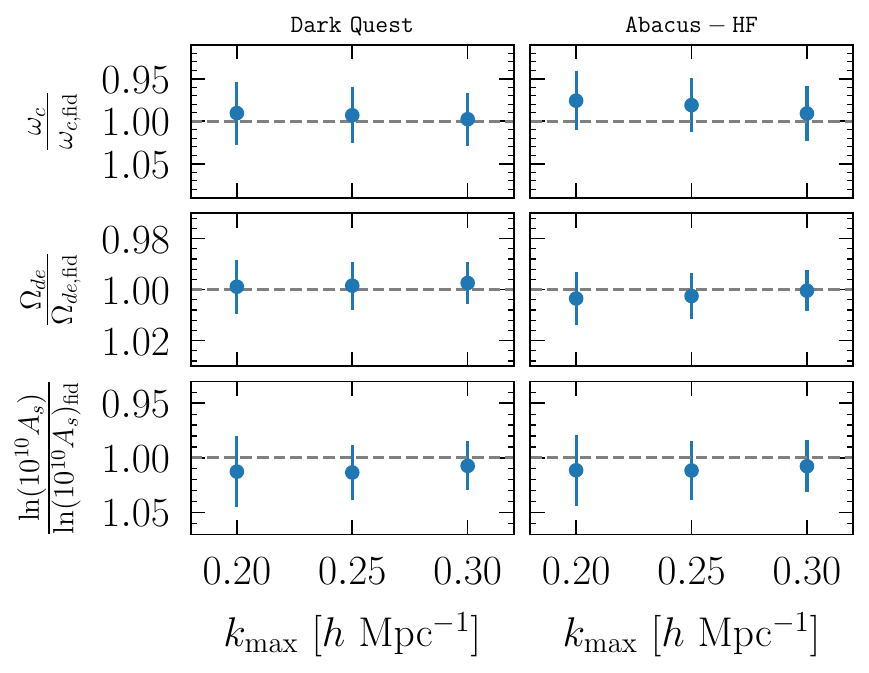}
    \includegraphics[width=0.45\textwidth]{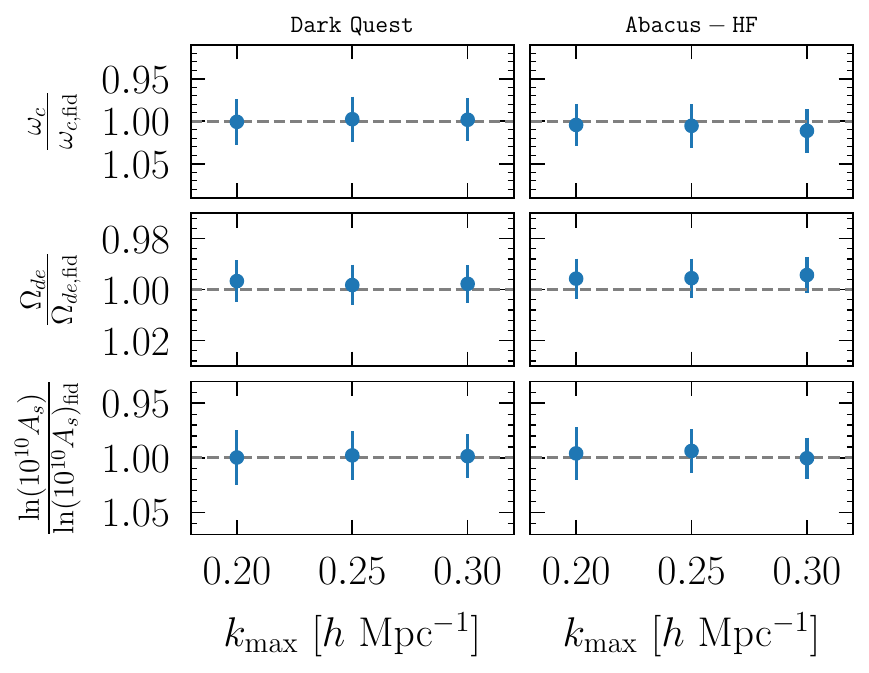}
    \caption{Dependence of the recovered cosmological parameters on the maximum wavenumber $k_{\rm max}$ from fits to $P_{\rm all}$ in the first redshift bin (left group) and in the second redshift bin (right group) under the $\Lambda$CDM model. Within each group, the left and right columns show results from the cubic {\tt Dark Quest} and {\tt Abacus-HF} mocks, respectively. The dashed horizontal line indicates the normalised fiducial value. The error bars denote the marginalized $1\,\sigma$ uncertainties, normalised by the fiducial values.}
    \label{fig:A11}
\end{figure*}

Fig.~\ref{fig:A11} shows the mock test results for different choices of $k_{\rm max}$. Overall, the cosmological parameters are well recovered as $k_{\rm max}$ increases. However, at $k_{\rm max} = 0.3 \,h\,\mathrm{Mpc}^{-1}$, the $\Omega_{de}$ constraint from the {\tt Abacus-HF} mocks in the higher-redshift bin lies at the edge of the $1\,\sigma$ interval relative to the fiducial value.

\begin{figure*}[htp]
    \centering
    \includegraphics[width=0.45\textwidth]{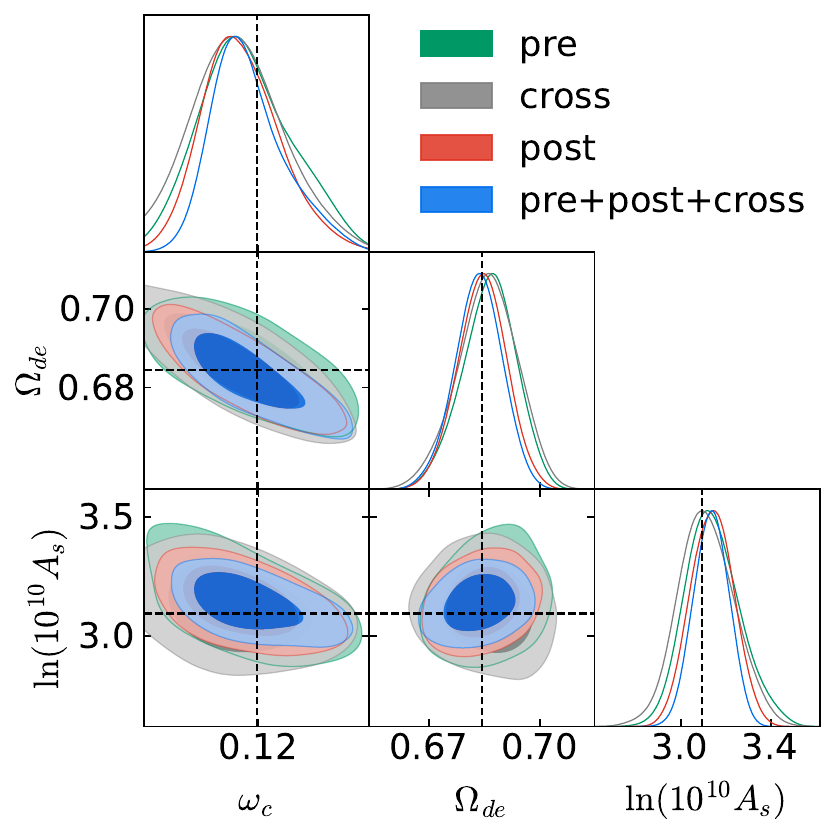}
    \includegraphics[width=0.45\textwidth]{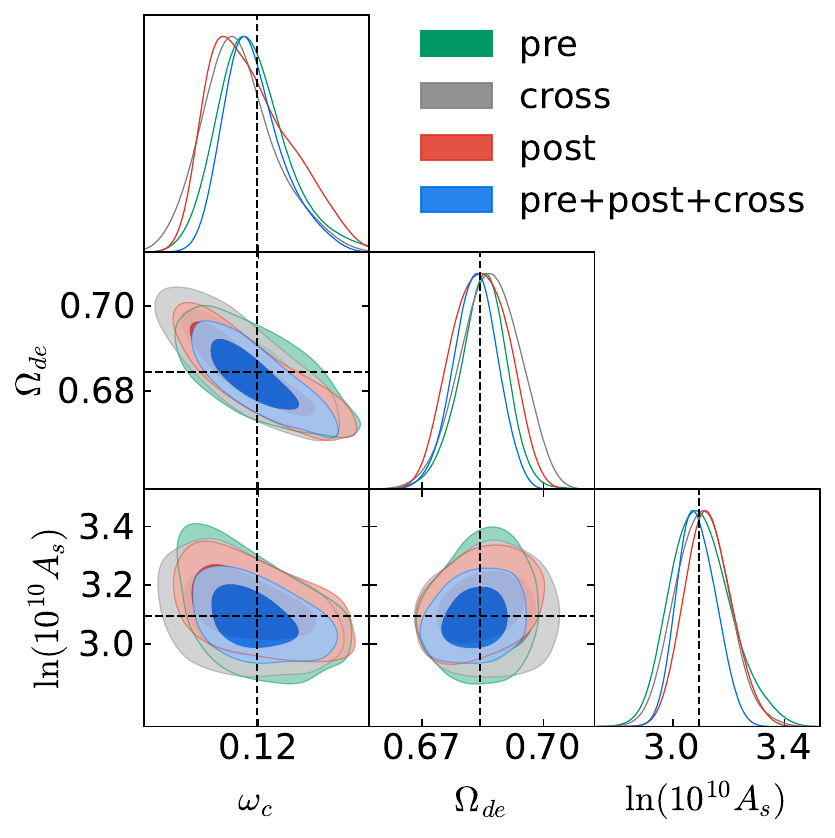}
    \caption{Marginalized posterior distributions of the cosmological parameters in the $\Lambda$CDM model, obtained from the average power-spectrum multipoles of the {\tt Dark Quest} fiducial mocks at redshifts $z=0.484$ (left panel) and $z=0.689$ (right panel). The dashed lines indicate the fiducial cosmology of the {\tt Dark Quest} simulations, and different colors correspond to fits using $P_{\rm pre}$, $P_{\rm post}$, $P_{\rm cross}$, and their combination $P_{\rm all}$.}
    \label{fig:A5}
\end{figure*}

\begin{figure*}[htp]
    \centering
    \includegraphics[width=0.45\textwidth]{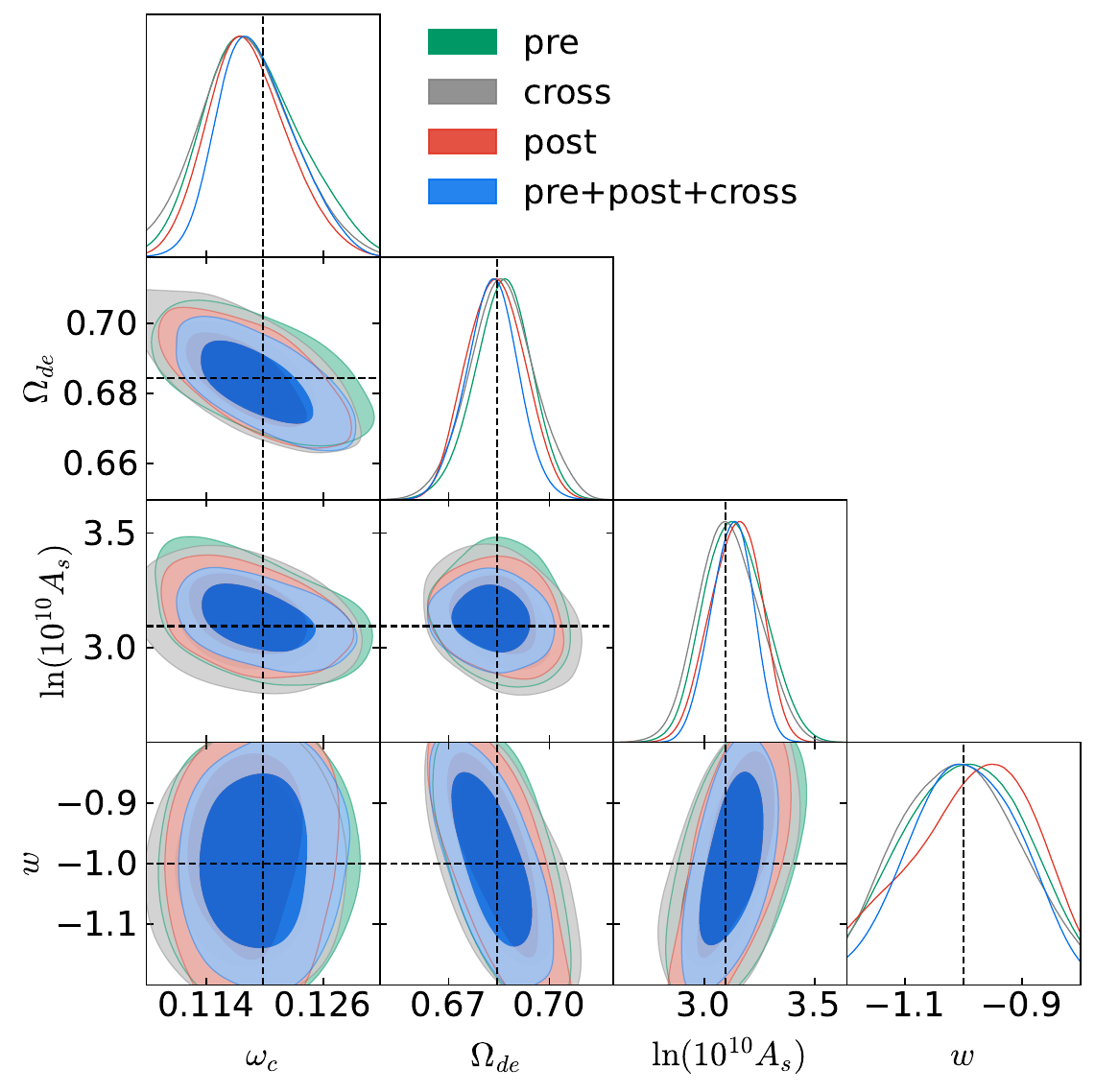}
    \includegraphics[width=0.45\textwidth]{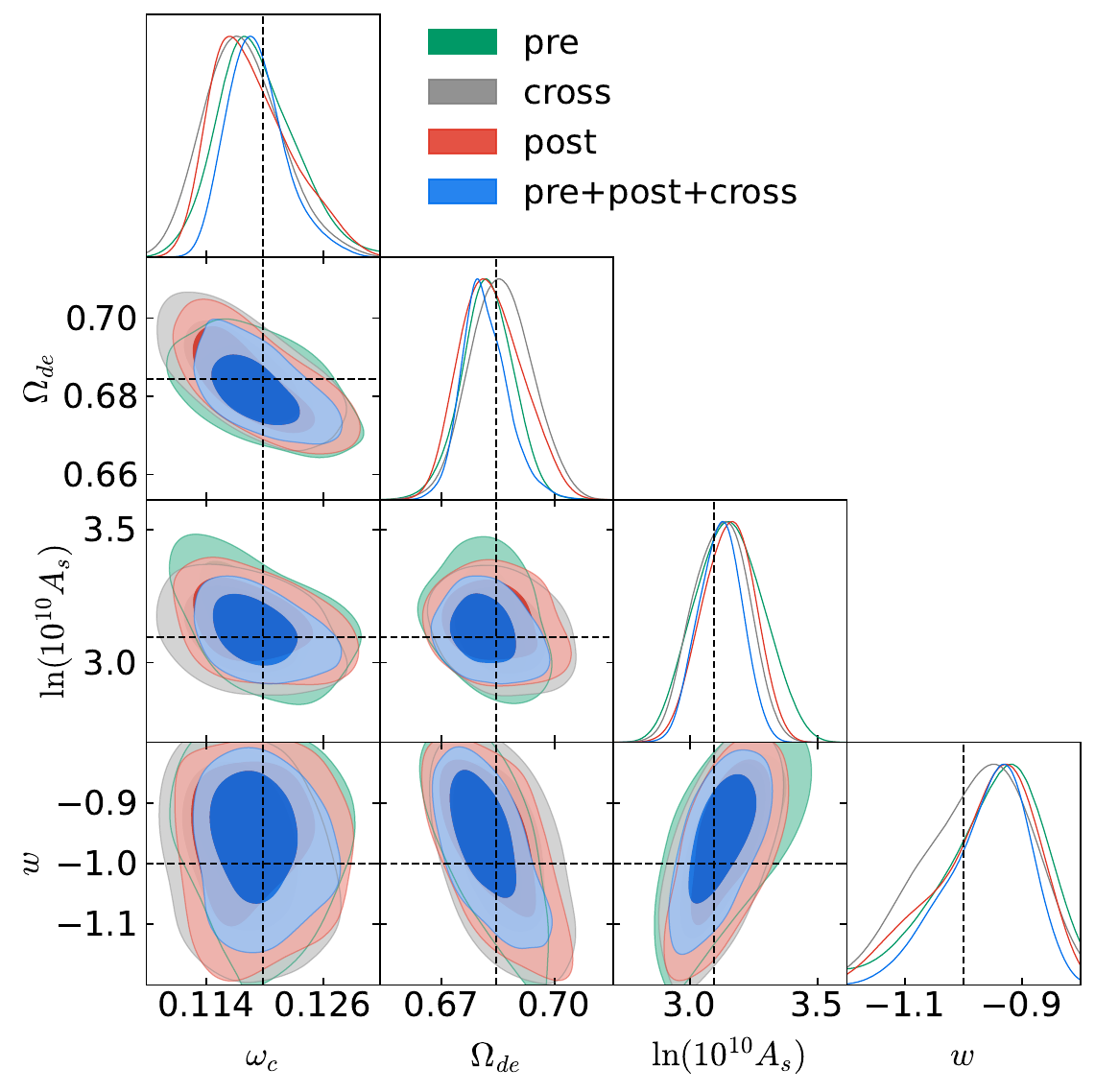}
    \caption{Same as Fig.~\ref{fig:A5}, but for the $w$CDM model in which the dark-energy EoS parameter $w$ is allowed to vary.}
    \label{fig:A6}
\end{figure*}

\begin{figure*}[htp]
    \centering
    \includegraphics[width=0.45\textwidth]{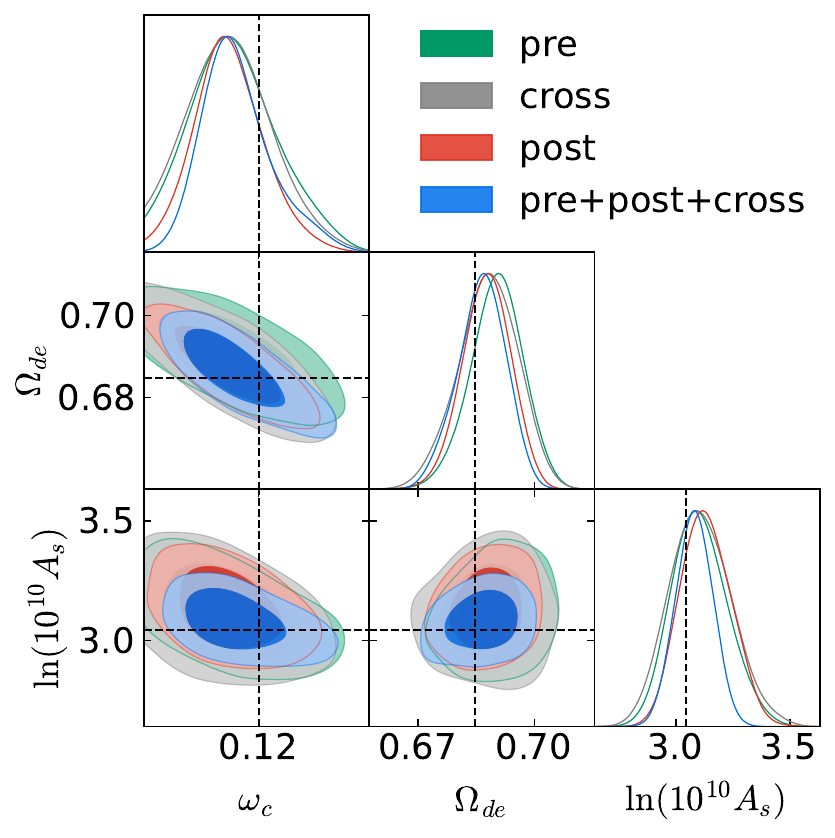}
    \includegraphics[width=0.45\textwidth]{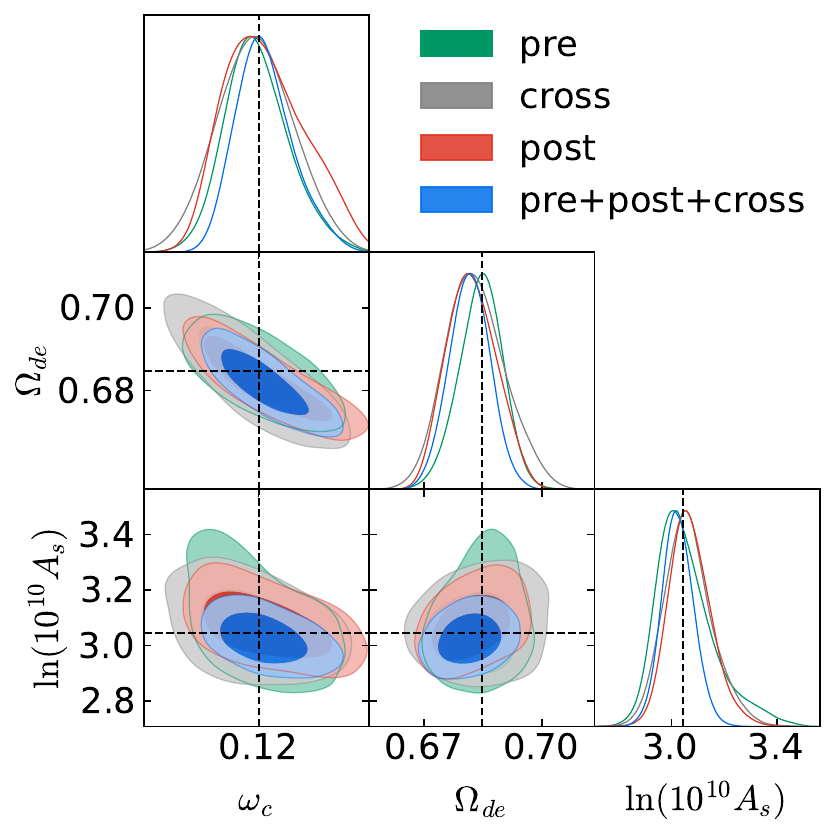}
    \caption{Marginalized posteriors on cosmological parameters in the $\Lambda$CDM model, obtained from the average power-spectrum multipoles of the {\tt Abacus-HF} fiducial mocks at redshifts $z=0.5$ (left panel) and $z=0.725$ (right panel). The dashed lines indicate the fiducial cosmology of the {\tt Abacus-HF} simulations.}
    \label{fig:A7}
\end{figure*}

\begin{figure*}[htp]
    \centering
    \includegraphics[width=0.45\textwidth]{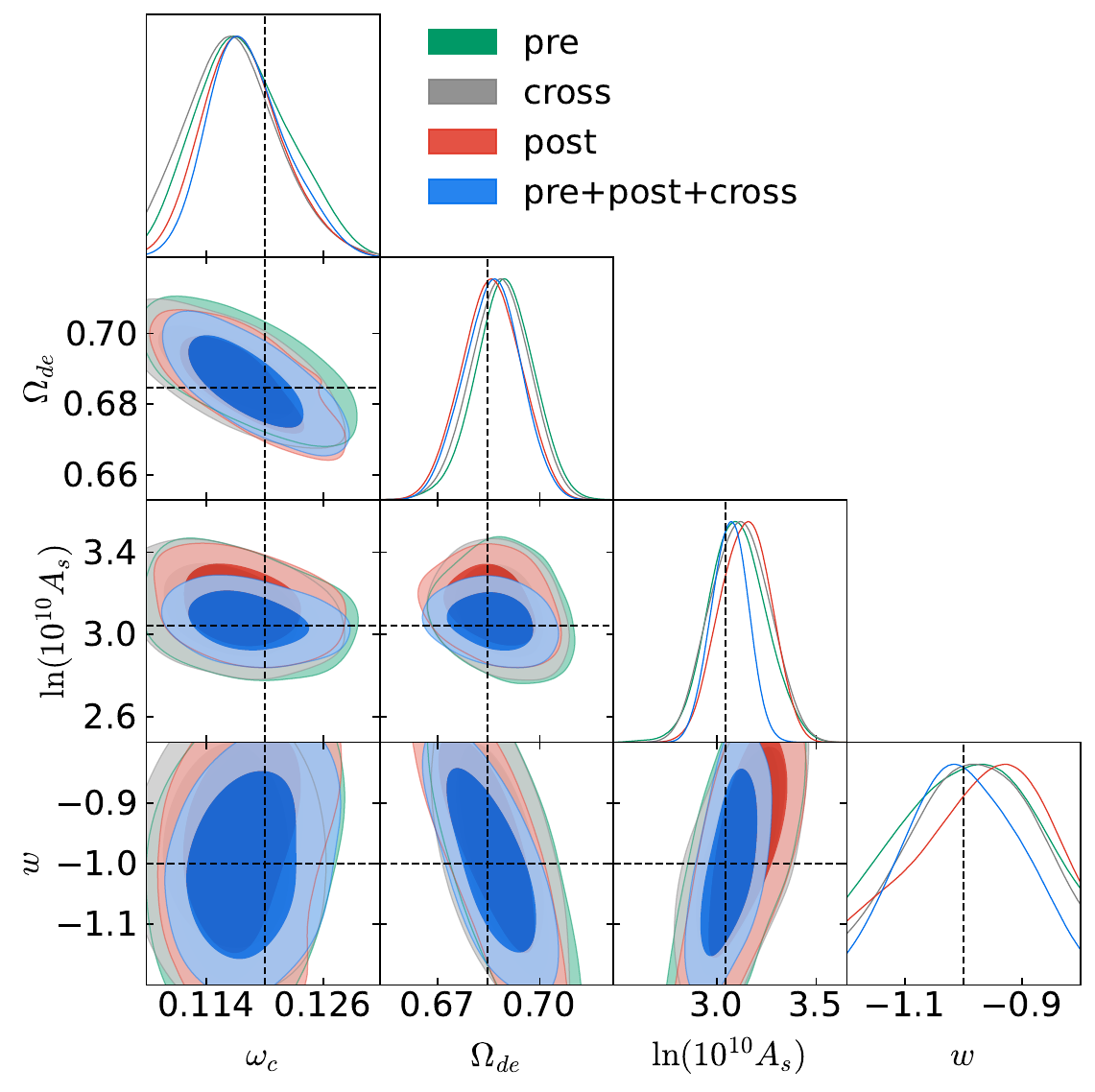}
    \includegraphics[width=0.45\textwidth]{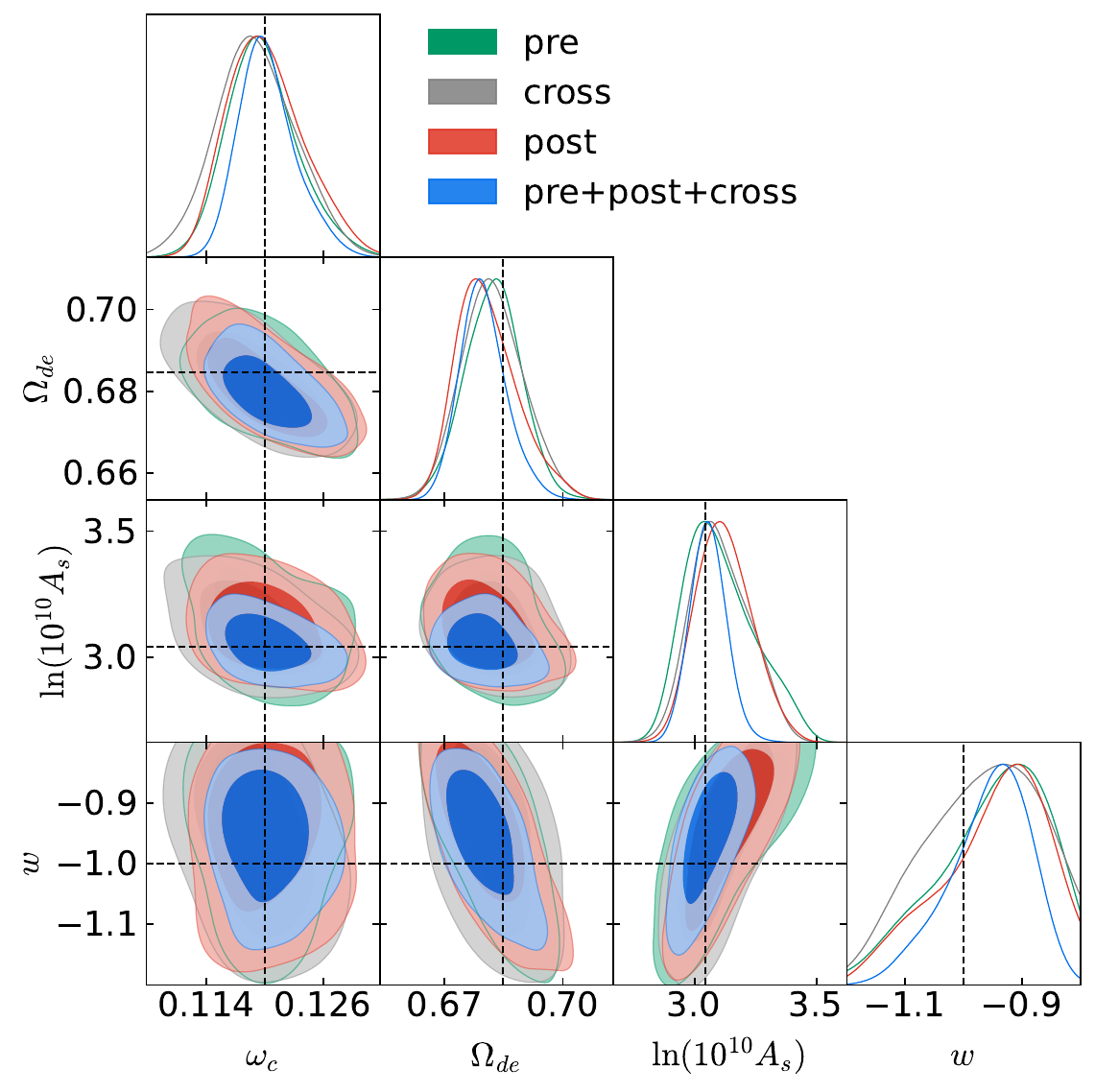}
    \caption{Same as Fig.~\ref{fig:A7}, but for the $w$CDM model with a free dark-energy EoS parameter $w$.}
    \label{fig:A8}
\end{figure*}

\begin{figure*}[htp]
    \centering
    \includegraphics[width=0.45\textwidth]{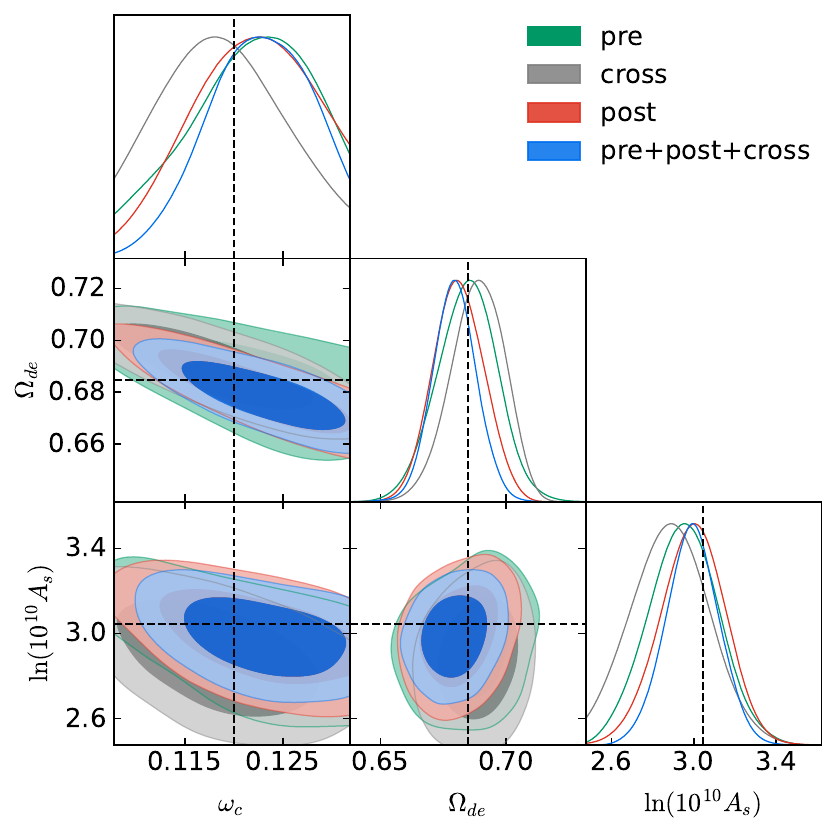}
    \includegraphics[width=0.45\textwidth]{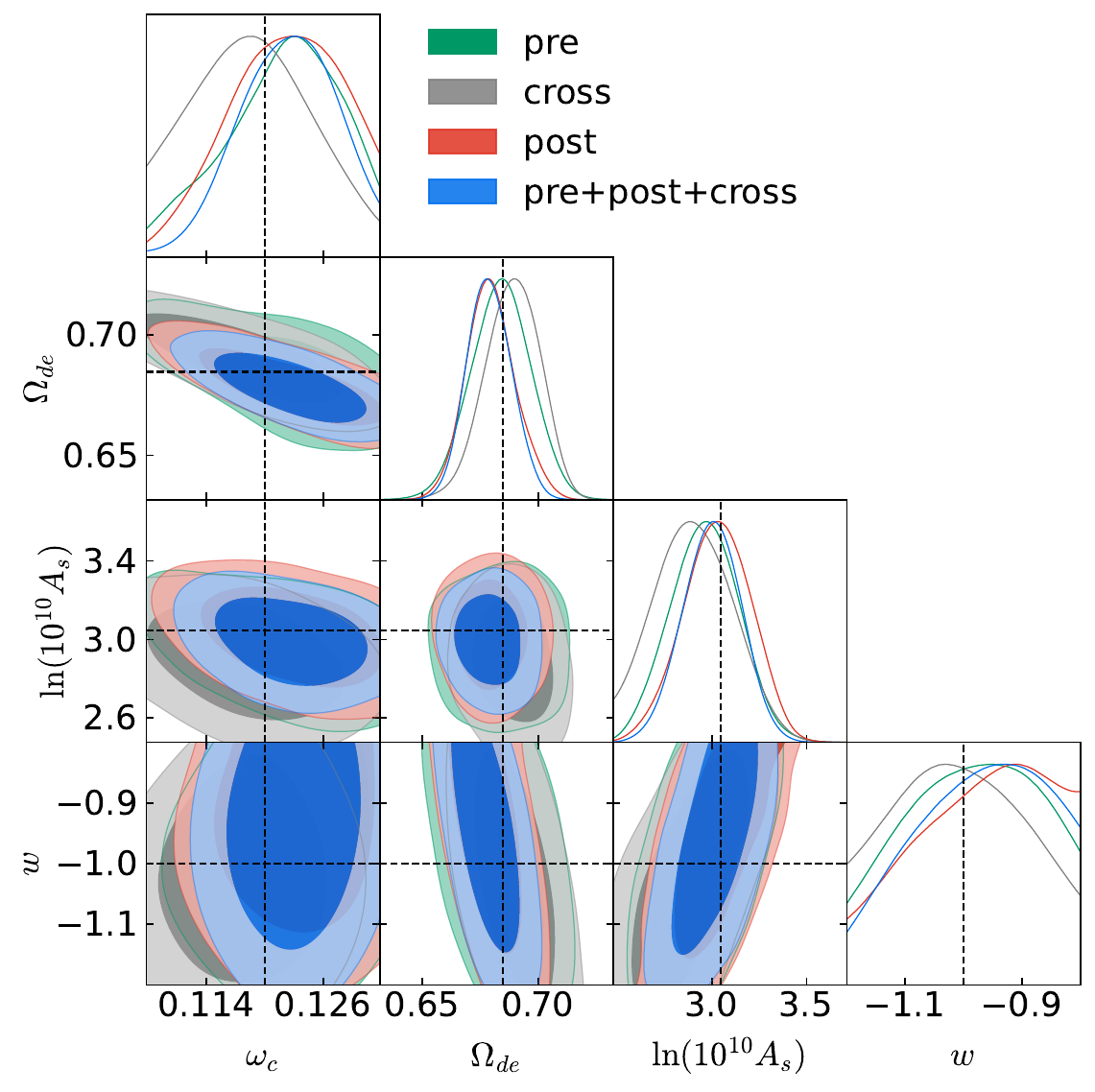}
    \caption{Marginalized posteriors on cosmological parameters in the $\Lambda$CDM (left) and $w$CDM (right) models, obtained from the average power-spectrum multipoles of the {\tt Abacus-2 complete} cut-sky mocks at redshift $z=0.5$. The agreement with the input cosmology indicates that the window-matrix implementation does not introduce noticeable biases.}
    \label{fig:A9}
\end{figure*}

In Figs.~\ref{fig:A5}-\ref{fig:A9}, we present the contour plots of mock tests based on different simulations.

\begin{figure*}[htp]
    \centering
    \includegraphics[width=0.9\textwidth]{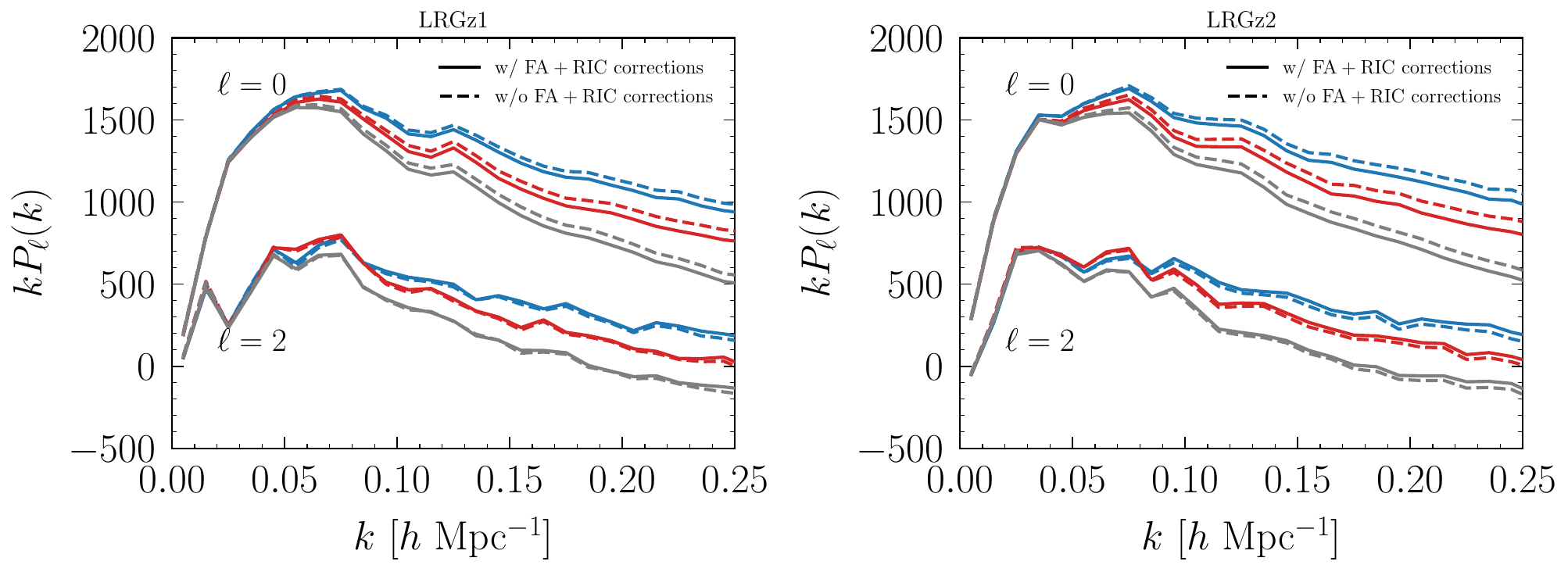}
    \caption{Left panel: monopole ($\ell=0$) and quadrupole ($\ell=2$) pre-reconstructed (blue lines), post-reconstructed (red lines), and cross power spectra (gray lines) of the DESI LRG sample in the first redshift bin. Solid lines show measurements including fibre-assignment (FA) incompleteness and radial integral constraint (RIC) corrections, while dashed lines denote results without these corrections. The comparison illustrates the impact of these two observational systematics on the measured power spectra across scales. Right panel: same as the right panel but for the second redshift bin.}
    \label{fig:A10}
\end{figure*}

Fig.~\ref{fig:A10} shows the comparison of the measured power spectra with and without FA+RIC corrections, visualizing the impact of these two observational systematics on the measured power spectra across scales.


The constraining results in $w$CDM from the combination of CMB distance priors with different choices of power-spectrum statistics $\rm P_{T}$ further augmented by three Type~Ia supernova datasets: Union3, PantheonPlus, and DES-Dovekie, are presented in Table \ref{tab:wCDM}.
 
\begin{table*}[!t]
\centering
\renewcommand{\arraystretch}{1.3}
\resizebox{\textwidth}{!}{%
\begin{tabular}{c|cccc|cccc}
\hline\hline
           & \multicolumn{4}{c|}{\texttt{LRG1} } & \multicolumn{4}{c}{\texttt{LRG2} } \\
\rule{0pt}{4ex}CMB+   &  $P_{\rm pre}$ & $ P_{\rm post}$ & $P_{\rm cross}$  & $P_{\rm all}$ &  $P_{\rm pre}$ & $ P_{\rm post}$ & $P_{\rm cross}$  & $P_{\rm all}$\\           
\rule{0pt}{4ex}$w$   &$-1.038 \pm 0.040 $&$-1.014 \pm 0.035 $&$-1.042 \pm 0.039 $&$-1.018 \pm 0.034 $&$-1.031 \pm 0.040 $&$-1.005 \pm 0.039$&$-1.024 \pm 0.045$&$-1.023 \pm 0.038$\\ 
$\Omega_m$  & $0.3032 \pm 0.0072$&$0.3107 \pm 0.0063$&$0.3016 \pm 0.0075 $&$0.3100 \pm 0.0061 $&$0.3087 \pm 0.0069 $&$0.3169 \pm 0.0060 $  &$0.3108 \pm 0.0086 $ &$0.3126 \pm 0.0069 $ \\ 
$H_0$  & $67.30 \pm 0.87 $ &$68.49 \pm 0.72$ &$69.47 \pm 0.88 $ &$68.59 \pm 0.70 $&$68.79 \pm 0.83 $&$67.94 \pm 0.73 $&$68.57 \pm 0.99 $&$68.43 \pm 0.78 $\\ 
$\sigma_8$  &$0.795 \pm 0.055 $&$ 0.840 \pm0.061$&$0.800 \pm 0.057 $&$0.800 \pm 0.040 $&$0.787 \pm 0.045 $&$0.782 \pm 0.044 $ &$0.774 \pm 0.052 $ &$0.763 \pm 0.037 $\\ 
\hline
\rule{0pt}{4ex}CMB+Union3+   &  $P_{\rm pre}$ & $ P_{\rm post}$ & $P_{\rm cross}$  & $P_{\rm all}$ &  $P_{\rm pre}$ & $ P_{\rm post}$ & $P_{\rm cross}$  & $P_{\rm all}$\\
\rule{0pt}{4ex}$w$  & $-0.994 \pm 0.030$ & $-0.980 \pm 0.028$ &$-0.995 \pm 0.030$& $-0.987 \pm 0.027$& $-0.985 \pm 0.031$&$-0.974 \pm 0.028$&$-0.985 \pm 0.029$&$-0.985 \pm 0.031$\\ 
$\Omega_m$   & $0.3105 \pm 0.0062$ & $0.3157 \pm 0.0054$ &$0.3102 \pm 0.0064$& $0.3144 \pm 0.0051$& $0.3149 \pm 0.0061$& $0.3200 \pm 0.0045$&$0.3169 \pm 0.0066$&$0.3183 \pm 0.0050$\\
$H_0$  & $68.31 \pm 0.65$& $67.79 \pm 0.57$ &$68.34 \pm 0.67$&$67.94 \pm 0.54$&$67.89 \pm 0.64$&$67.41 \pm 0.49$&$67.59 \pm 0.67$&$67.63 \pm 0.54$\\
$\sigma_8$    & $0.800\pm0.057$ & $0.850 \pm 0.066$ &$0.818 \pm 0.059$&$0.798 \pm 0.044$&$0.798 \pm 0.051$&$0.785 \pm 0.044$&$0.795 \pm 0.053$&$0.769 \pm 0.038$\\ 
\hline
\rule{0pt}{4ex}CMB+PantheonPlus+   &  $P_{\rm pre}$ & $ P_{\rm post}$ & $P_{\rm cross}$  & $P_{\rm all}$ &  $P_{\rm pre}$ & $ P_{\rm post}$ & $P_{\rm cross}$  & $P_{\rm all}$\\
\rule{0pt}{4ex}$w$  &$-0.995 \pm 0.025$ & $-0.987 \pm 0.025$ &$-0.993 \pm 0.025$&$-0.992 \pm 0.024$&$-0.992 \pm 0.026$&$-0.984 \pm 0.027$&$-0.987 \pm 0.028$&$-0.993 \pm 0.027$\\ 
$\Omega_m$   & $0.3103 \pm 0.0053$ & $0.3146 \pm 0.0049$ &$0.3104 \pm 0.0062$&$0.3135 \pm 0.0049$&$0.3140 \pm 0.0053$& $0.3195 \pm 0.0045$&$0.3166 \pm 0.0058$&$0.3170 \pm 0.0051$\\
$H_0$  & $68.33 \pm 0.53$& $67.93 \pm 0.49$&$68.31 \pm 0.61$&$68.07. \pm 0.48$& $68.02 \pm 0.53$&$67.55 \pm 0.47$&$67.78 \pm 0.56$&$67.81 \pm 0.51$\\
$\sigma_8$    & $0.800\pm0.055$ & $0.844 \pm 0.061$ &$0.821 \pm 0.060$&$0.798 \pm 0.045$&$0.792 \pm 0.048$&$0.785 \pm 0.045$&$0.788 \pm 0.054$&$0.767 \pm 0.036$\\ 
\hline
\rule{0pt}{4ex}CMB+DES-Dovekie+ &  $P_{\rm pre}$ & $ P_{\rm post}$ & $P_{\rm cross}$  & $P_{\rm all}$ &  $P_{\rm pre}$ & $ P_{\rm post}$ & $P_{\rm cross}$  & $P_{\rm all}$\\
\rule{0pt}{4ex}$w$  &$ -0.990 \pm 0.025$ &$-0.981 \pm 0.023$ &$-0.990 \pm 0.026 $&$ -0.988 \pm 0.023$&$-0.987 \pm 0.025 $&$-0.983 \pm 0.024 $&$ -0.984 \pm 0.027$&$-0.988 \pm 0.025 $\\ 
$\Omega_m$   &$0.3111 \pm 0.0053 $ &$ 0.3149 \pm 0.0048 $ &$0.3111 \pm 0.0057 $&$0.3141 \pm 0.0048 $&$0.3144 \pm 0.0051 $&$0.3196 \pm 0.0042 $&$0.3170 \pm 0.0057 $&$0.3179 \pm 0.0046 $\\
$H_0$  &$68.22 \pm 0.51 $ &$67.86 \pm 0.46 $ &$ 68.22 \pm 0.56$&$67.98 \pm 0.45 $&$67.94 \pm 0.50 $&$67.54 \pm 0.42 $&$67.73 \pm 0.53 $&$ 67.69 \pm 0.45$\\
$\sigma_8$    &$0.800 \pm 0.056 $ &$0.846 \pm 0.063 $ &$0.824 \pm 0.057 $&$0.799 \pm 0.042 $&$0.794 \pm 0.049 $&$0.785 \pm 0.048 $&$0.790 \pm 0.053 $&$0.775 \pm 0.042 $\\
\hline\hline
\end{tabular}
}
\renewcommand{\arraystretch}{1}
\caption{Summary of constraints on the cosmological parameters ($w,\Omega_m, H_0,\sigma_8$) in $w$CDM model from CMB+$P_{\rm T}$ without supernovae, and from CMB+$P_{\rm T}$ further combined with the Union3, PantheonPlus, and DES-Dovekie SN~Ia datasets, respectively.}
\label{tab:wCDM}
\end{table*}

\end{document}